\documentclass[nohyper]{JHEP3}
\usepackage{amsmath,amssymb}
\usepackage{graphicx}
\usepackage{cite}
\usepackage{ifthen}
\newcommand{\Sla}[1]{/\!\!\!\!#1}
\def\ie{{\it i.e. }}
\def\lsim{\raise0.3ex\hbox{$\;<$\kern-0.75em\raise-1.1ex\hbox{$\sim\;$}}}
\def\gsim{\raise0.3ex\hbox{$\;>$\kern-0.75em\raise-1.1ex\hbox{$\sim\;$}}}
\title{Neutrino Masses at LHC: 
Minimal Lepton Flavour Violation in Type-III See--saw}
\author{O.\ J.\ P.\ \'Eboli\\
Instituto de F\'{\i}sica,
Universidade de S\~ao Paulo, S\~ao Paulo -- SP, Brazil.\\
E-mail: \email{eboli@fma.if.usp.br}}
\author{J.\ Gonzalez-Fraile\\
  Departament d'Estructura i Constituents de la Mat\`eria and
  ICC-UB, Universitat de Barcelona, 647 Diagonal, E-08028 Barcelona,
  Spain\\
E-mail: \email{fraile@ecm.ub.es}}
\author{M.\ C.\ Gonzalez--Garcia\\
  Instituci\'o Catalana de Recerca i Estudis Avan\c{c}ats (ICREA),
  Departament d'Estructura i Constituents de la Mat\`eria, Universitat
  de Barcelona, 647 Diagonal, E-08028 Barcelona, Spain\\
{\rm and:}
  C.N.~Yang Institute for Theoretical Physics, SUNY at Stony Brook,
  Stony Brook, NY 11794-3840, USA\\
E-mail: \email{concha@insti.physics.sunysb.edu}}

\abstract{
  We study the signatures of minimal lepton flavour violation in a
  simple Type-III see--saw model in which the flavour scale is given
  by the new fermion triplet mass and it can be naturally light enough
  to be produced at the LHC. In this model the flavour structure of
  the lepton number conserving couplings of the triplet fermions to the 
  Standard Model leptons
  can be reconstructed from the neutrino mass matrix and the smallness
  of the neutrino mass is associated with a tiny violation of total
  lepton number.  Characteristic signatures of this model include
  suppressed lepton number violation decays of the triplet fermions,
  absence of displaced vertices in their decays and predictable lepton
  flavour composition of the states produced in their decays. We study
  the observability of these signals in the processes $pp\rightarrow
  3\ell + 2j +\Sla{E_T}$ and $pp\rightarrow 2\ell + 4j$ with $\ell =e$
  or $\mu$ taking into account the present low energy data on neutrino
  physics and the corresponding Standard Model backgrounds.  Our
  results indicate that the new fermionic states can be observed for
  masses up to 500 GeV depending on the CP violating Majorana phase
  for an integrated luminosity of 30 fb$^{-1}$. Moreover, the flavour
  of the final state leptons in the above processes can shed light on
  the neutrino mass ordering.}
\preprint{YITP-SB-11-28\\ICCUB-11-163}

\begin{document}

\section{Introduction}

The observation of neutrino masses and mixing is our first evidence of
physics beyond the Standard Model (SM) ~\cite{review}.  The CERN Large
Hadron Collider (LHC) has started operation with the aim of exploring
physics beyond SM at the TeV scale.  Then, an obviously interesting
question is whether the new physics (NP) associated to the neutrino
masses can be within the LHC reach. Here we analyze the LHC potential
to unravel the existence of triplet fermionic states that appear in
minimal flavour violating theories based on Type-III see--saw
models of neutrino mass.
We also discuss how to probe the neutrino mass ordering in the
production and decay of these new possible states.

NP effects at energies below the characteristic NP scale can be
described in terms of effective higher-dimension ($d>4$) operators and
it is well known that, with the particle contents of the SM, there is
just one dimension-five operator which can be built~\cite{Weinberg},
$\alpha_5/\Lambda_{\rm LN} {\cal O}_5=\alpha_5/\Lambda_{\rm LN} L_L^{}
L_L^{} HH$, where $L_L^{}$ and $H$ are the leptonic and Higgs
$SU(2)_L$ doublets.  This operator breaks total lepton number and
after electroweak symmetry breaking it generates Majorana masses for
the neutrinos $m_\nu \sim \alpha_5 v^2/\Lambda_{LN}$, where $v$ is the
SM Higgs vacuum expectation value (vev).  Consequently neutrinos are
much lighter than the other SM fermions because of the large scale of
total lepton number violation $\Lambda_{LN}$.  In simple
renormalizable realizations of NP this dimension-5 operator can be
generated by the tree-level exchange of three types of new particles:
\begin{itemize}
\item Type-I see--saw~\cite{TypeI}: One adds at least two fermionic
  singlets of mass $M$ and the neutrino masses are $m_\nu \sim
  \lambda^2 v^2/ M$, where $\lambda$ is the Yukawa coupling.
\item Type-II see--saw~\cite{TypeII}: One adds an $SU(2)_L$ Higgs
  triplet $\Delta$ of mass $M$ and with a neutral component which in
  presence of scalar doublet-triplet mixing $\mu$ term in the scalar
  potential acquires a vev $v_\Delta=\mu v^2/M^2$.  The neutrino
  masses are $m_\nu \sim \lambda {\mu v^2}/{M^2}$.
\item Type-III see--saw~\cite{TypeIII}: One adds at least two
  $SU(2)_L$ fermion triplets with zero hypercharge generating neutrino
  masses, $m_\nu \sim \lambda^2 v^2/M$.
\end{itemize}
In addition hybrid scenarios containing some combination of these
states have also been constructed \cite{hybrid}.  In any of these
mechanisms the smallness of the neutrino mass can be naturally
explained with Yukawa couplings $\lambda\sim{\cal O}(1)$ if the masses
of the heavy states are $M\sim \Lambda_{LN}\sim 10^{14-15}$ GeV (with
$\mu\sim M$ also for Type-II), clearly out of reach of the LHC.

Since strictly nothing prevents that the new states have TeV scale
masses, there has been an increasing literature studying the possible
signatures of these neutrino--mass--inducing states with TeV-scale
masses at the
LHC (see e.g. ~\cite{delAguila:2008cj,delAguila:2008hw,han,strumia,li,bajc}).
Nevertheless, in some cases such a low-scale $M$ is technically
unnatural or, in some others, it is simply not very well motivated
theoretically. Notwithstanding consistent models of TeV-scale see--saw
exist in the literature for some time (see e.g. \cite{inverse,smirnov}).

Generically, at low energies the Lagrangian of the full theory can be
expanded as
\begin{equation}
{\cal L}={\cal L}_{SM}+\frac{\alpha_5}{\Lambda_{LN}} {\cal O}_5 +
\sum_i \frac{\alpha_{6,i}}{\Lambda^2_{FL}} 
{\cal O}_{6,i} 
+\dots
\label{eq:efflag}
\end{equation}
where ${\cal O}_5 $ is Weinberg's operator responsible for neutrino
masses, and ${\cal O}_{6,i}$ are flavour-changing, but lepton number
conserving, dimension-6 operators. In this context attractive
TeV-scale see--saw models are those for which it is possible to relate
the mass of the new states $M\sim \Lambda_{FL}\sim {\cal O}$ (TeV) but
still keep $\Lambda_{LN}\gg \Lambda_{FL}$ to explain the smallness of
the neutrino mass. This is different than the simplest
implementations described above for which $M \sim\Lambda_{LN} \sim
\Lambda_{FL}$.

In this effective operator approach the possibility of TeV scale
see--saw models has been recently revised in the context of minimal
lepton flavour violation (MLFV)
~\cite{cgiw,davidson,Gavela:2009cd,alonso}. Minimal flavour violation was
first introduced for quarks \cite{mfv} as a way to explain the absence
of NP effects in flavour changing processes in meson decays. The basic
assumption is that the only source of flavour mixing in the full
theory is the same as in the SM, \ie the quark Yukawa couplings.  This
idea was latter on extended for leptons \cite{cgiw,davidson} although
for leptons the precise hypothesis corresponding to MLFV is less
well-defined as the SM only contains Yukawa couplings for the charged
leptons and those are not enough to explain the neutrino
data. Consequently, the couplings and generation structure of the new
states must also be considered when defining the conditions for MLFV,
making them model dependent by default.

In Ref.~\cite{Gavela:2009cd} simple see--saw models were constructed
which realize the conditions associated with MLFV as set-up in
Ref.~\cite{cgiw}, {\em i.e.} there is large hierarchy between the
lepton number and lepton flavour breaking scales,
$\Lambda_{LN}\gg\Lambda_{FL}$, and the coefficients $\alpha_{6,i}$ are
determined by $\alpha_5$. As discussed in Ref.~\cite{Gavela:2009cd}
these conditions are automatically fulfilled by the simplest Type-II
see--saw model if a light double-triplet mixing $\mu$ is assumed.  For
LHC phenomenology this leads to the interesting possibility studied in
detail in Refs.~\cite{han,julia} of the production of the triplet
scalar states with all their decay modes determined by the neutrino
mass parameters. From the theoretical side, one drawback of such a
scenario is that it is difficult to keep such a low $\mu$ stable 
if generated by spontaneous breaking of lepton number.
In the same work~\cite{Gavela:2009cd} the authors presented a very
simple model for Type-I or Type-III see--saw with naturally light
states. From the point of view of LHC phenomenology these models are
attractive since, a) the new states can be light enough to be produced
at LHC, and b) their observable (ie the lepton number conserving) 
signatures are fully determined by the neutrino
parameters. In Type-I see--saw the new states are SM singlets, and
therefore, they can only be produced via their mixing with the SM
neutrinos. This leads to small production rates which makes the model
only marginally testable at LHC. Type-III see--saw fermions, on the
contrary, are SM triplets with weak-interaction pair-production cross
section, and consequently, having the potential to allow for tests of
the hypothesis of MLFV. This is the scenario which we explore in this
work. Alternatively, signals of MLFV in a model in which the MLFV
condition does not involve the states associated with the generation
of neutrino mass has been explored in Ref.~\cite{nir}.

The outline of this work is as follows. In
Sec.~\ref{sec:model} we summarize the basics of the model in which the
flavour scale is given by the new fermion triplet mass and it can be
naturally light to be within reach at LHC. We describe how in this
model the flavour structure of the lepton number conserving 
couplings of the triplet fermions,
and consequently their observable decay branching ratios to the 
SM leptons, can
be reconstructed from the neutrino mass matrix. In this model the
lightness of the neutrino mass implies that lepton number violating
decay modes of the triplet fermions are suppressed. Section
\ref{sec:sig} describes the generic features of the expected total
lepton number conserving signatures. In Sec.~\ref{sec:llljjn} we
evaluate in detail the signal and backgrounds for the process
$pp\rightarrow\ell\ell\ell jj \Sla{E_T}$ for which the challenge is
the identification/assignment of the lepton corresponding to the decay
chain of each of the fermion triplets. Section~\ref{sec:lljjjj}
contains our analysis of the discovery potential of the process
$pp\rightarrow\ell\ell jjjj$ which presents a larger QCD
background. For both final states we evaluate the expected signals
within the presently allowed ranges of neutrino parameters, we study
their statistical significance as a function of those for fermion
triplet masses in the range $150$--$500$ GeV and we also discuss how
to probe the neutrino mass ordering in these final states. These
studies are done for the LHC running at 14 TeV. We comment on
Sec.~\ref{sec:7tev} the potential of the 7 TeV run.  Finally we
summarize our conclusions in Sec.~\ref{sec:conclu}.

\section{Simplest MLFV Type--III see--saw model}
\label{sec:model}

We describe here the simplest MLFV model presented in
Ref.~\cite{Gavela:2009cd} adapted for Type-III see--saw. As explained
above, Type-I see--saw heavy fermions are SM singlets, and therefore,
they can only be produced via their mixing with the SM neutrinos. This
leads to small production rates which makes the model only marginally
testable at LHC.

In this MLFV Type-III see--saw model the SM Lagrangian is extended
with two fermion triplets $\vec{\Sigma} = \left(
  \Sigma_1,\Sigma_2,\Sigma_3\right) $ and $\vec{\Sigma}^\prime =
\left( \Sigma'_1,\Sigma'_2,\Sigma'_3\right)$, each one formed by three
right-handed Weyl spinors of zero hypercharge. Hence, the Lagrangian
is
\begin{eqnarray}
 {\mathcal L}={\mathcal L}_{SM}+{\mathcal L}_K+{\mathcal
 L}_Y+{\mathcal L}_\Lambda
\label{eq:lag}
\end{eqnarray}
with
\begin{eqnarray}
{\mathcal
L}_K=&&i\left(\overline{\vec{\Sigma}}\Sla{D}_\mu\vec{\Sigma}+
\overline{\vec{\Sigma}^\prime}\Sla{D}_
\mu\vec{\Sigma}^\prime\right)\\ {\mathcal
L}_Y=&&-Y_i^\dag\overline{L^w_{L
i}}\left(\vec{\Sigma}\cdot\vec{\tau}\right)\tilde{\phi}- \epsilon
Y_i^{\prime\dag}\overline{L^w_{L
i}}\left(\vec{\Sigma}^\prime\cdot\vec{\tau}\right)\tilde{\phi}+h.c.\\
{\mathcal
L}_\Lambda=&&
-\frac{\Lambda}{2}\left(\overline{\vec{\Sigma}^c}\vec{\Sigma}^\prime
+\overline{\vec{\Sigma}^{\prime c}}\vec{\Sigma}\right)
-\frac{\mu}{2}\overline{\vec{\Sigma}^{\prime c}}\vec{\Sigma}^\prime
-\frac{\mu'}{2}\overline{\vec{\Sigma}^c}\vec{\Sigma}
+h.c.
\end{eqnarray}
Here $\vec{\tau}$ are the Pauli matrices, the gauge covariant
derivative is given by $D_\mu=\partial_\mu+ig\vec{T}\cdot\vec{W}_\mu$,
where $\vec{T}$ are the three-dimensional representation of the
$SU(2)$ generators, $\phi$ is the SM Higgs doublet, and
$L^w_{i}=(\nu^w_i,\ell^w_i)^T$ are the three lepton doublets of the
SM. The index $w$ makes reference at the fact that these are
weak-eigenstates to be distinguish from those without the index which
will be the mass eigenstates.

In the MLFV Type--III see-saw model the flavour-blind parameters
$\epsilon$, $\mu$ and $\mu'$ are {\sl small}, ie, the scales $\mu$ and
$\mu'$ are much smaller than $\Lambda$ and $v$ and $\epsilon\ll 1$.
As a consequence the Lagrangian in Eq.~\eqref{eq:lag} breaks total
lepton number due to the simultaneous presence of the Yukawa terms
$Y_i$ and $\epsilon Y'_i$ as well as to the presence of the $\mu$ and
$\mu'$ terms.  In the limit $\mu,\mu',\epsilon\rightarrow 0$ it is
possible to define a conserved total lepton number by assigning
$L(L^w)=L(\Sigma)=-L(\Sigma^\prime)=1$.

After electroweak symmetry breaking and working in the unitary gauge,
$\tilde{\phi}^T=\frac{1}{\sqrt{2}}\left(v\ \ \ 0\right)$, the six Weyl
fermions of well defined electric charge are
$\Sigma^{(\prime)}_\pm=\frac{1}{\sqrt{2}}\left(\Sigma^{(\prime)}_1\mp
  i\Sigma^{(\prime)}_2\right)$ and
$\Sigma^{(\prime)}_0=\Sigma^{(\prime)}_3$. From those one defines the
negatively charged Dirac fermions $E$ and $E'$ and the neutral
Majorana fermions $\tilde N$ and $\tilde N^\prime$
\begin{eqnarray}
E^{(\prime)}
=\Sigma^{(\prime)}_-+{\Sigma_+^{(\prime)}}^c &\;\;\;\; 
\tilde N^{(\prime)}=\Sigma^{(\prime)}_0+{\Sigma^{(\prime)}_0}^c 
\; .
\end{eqnarray}
In this basis the leptonic mass terms read
\begin{eqnarray}
{\mathcal L}_m
&=&-\frac{1}{2}\left(\overline{\vec{\nu^w_L}^c}\ 
\overline{\tilde N_R}\ \overline{\tilde N_R^{\prime}}\right)
M_0\left(\begin{array}{c}
       \vec{\nu^w_L}\\\tilde N_R^c\\\tilde N_R^{\prime c}
      \end{array}\right)
-\left(\overline{\vec{\ell^w_L}}\ \overline{E_L}\ 
\overline{E_L^{\prime}}\right)
M_\pm
\left(\begin{array}{c}\vec{\ell^w_R}\\E_R\\E_R^{\prime}\end{array}\right)
+h.c
\label{eq:lmass}
\end{eqnarray}
with
\begin{eqnarray}
M_0=\left(\begin{array}{ccc}
       0&\frac{v}{\sqrt{2}}Y^T&\epsilon\frac{v}{\sqrt{2}}Y^{\prime T} \\
  \frac{v}{\sqrt{2}}Y&\mu'&\Lambda \\
\epsilon\frac{v}{\sqrt{2}}Y^{\prime}&\Lambda&\mu
      \end{array}\right) 
&&
M_\pm=
\left(\begin{array}{ccc}
\frac{v}{\sqrt{2}}Y^\ell&vY^\dagger&\epsilon vY^{\prime \dagger} \\
0&\mu'&\Lambda \\
0&\Lambda&\mu
\end{array}\right)
\end{eqnarray}
where $Y^\ell$ are the charged lepton Yukawa couplings of the SM and
$Y^{(\prime)}=(Y^{(\prime)}_1,Y^{(\prime)}_2,Y^{(\prime)}_3)$.  In
writing Eq.~(\ref{eq:lmass}) we denote by $\vec\nu^w$ and $\vec\ell^w$
two column vectors containing the three neutrinos and charged leptons
of the SM in the weak basis respectively. Furthermore, without loss of
generality, we have chosen to work in a basis in which $\Lambda$ is
real while both $Y$ and $Y'$ are complex. In general the parameters $\mu$ and
$\mu'$ would be complex, but for the sake of simplicity we 
have taken them to be real in what follows though it is  
straight forward to generalize the expression  to include the relevant phases
~\cite{Abada:2007ux}.

Diagonalizing ${\mathcal L}_m$ one finds three light Majorana
neutrinos $\nu_i$ -- the lightest being massless -- and three light
charged massive leptons $\ell_i$ that satisfy
\begin{eqnarray}
m^{diag}_\nu&=& 
{V^\nu}^T \left[-\frac{v^2}{2\Lambda} 
\epsilon\left({\widehat{Y'}}^T Y+ Y^T {\widehat {Y'}}\right)\right]V^\nu  \; , \\
m^{diag}_\ell&=&
\frac{v}{\sqrt{2}}  {V^\ell}^\dagger_R Y^{\ell\dagger}
\left[1-\frac{v^2}{2\Lambda^2} 
Y^\dagger Y\right] V^\ell_L \; .
\end{eqnarray}
$V^\nu$ and $V^\ell_{L,R}$ being $3\times 3$ unitary matrices
and for convenience we have defined the combination
\begin{equation}
\widehat {Y'}=Y'-\frac{1}{\epsilon}\frac{\mu}{2\Lambda} Y  \: .
\label{eq:haty}
\end{equation}
One finds also two heavy Majorana neutral leptons and two charged
heavy leptons with masses $M\simeq\Lambda (1 \mp
\frac{\mu+\mu'}{2\Lambda})$.  We construct a quasi-Dirac state $N$
with the two Majorana neutral leptons and two combinations of the
heavy charged leptons $E^-_{1}$ and $E^+_2$ which are related to the
weak eigenstates by
\begin{eqnarray}
\nu_L^w&=&V^\nu \nu_L 
+ \frac{v}{\sqrt{2}\Lambda}Y^\dagger N_L
+ \frac{v}{\sqrt{2}\Lambda}\left(\epsilon Y^{\prime \dagger}-
\left(\frac{3\mu+\mu^\prime}{4\Lambda}\right)Y^\dagger\right)N_R^c
\;, 
\\
\ell^w_L&=&\ell_L 
+\frac{v}{\Lambda}Y^\dagger E_{1L}^-
+\frac{v}{\Lambda}\left(\epsilon Y^{\prime \dagger}-\left(\frac{3\mu+\mu^\prime}{4\Lambda}\right)
Y^\dagger\right)E_{2R}^{+c}  \; , \\
\ell^w_R&=&\ell_R \; , \\
N_L&=&N_R^c-\left(\frac{\mu-\mu^\prime}{4\Lambda}\right)N_L
-\frac{v}{\sqrt{2}\Lambda}\left(\epsilon Y^{\prime}-\frac{\mu}{\Lambda}Y\right)V^\nu \nu_L\; , \\
N'_L&=&N_L+\left(\frac{\mu-\mu^\prime}{4\Lambda}\right)N_R^c
-\frac{v}{\sqrt{2}\Lambda}Y V^\nu \nu_L    \;,
\\
E_L&=&E_{2R}^{+c}
-\left(\frac{\mu-\mu^\prime}{4\Lambda}\right)E_{1L}^-
-\frac{v}{\Lambda}\left(\epsilon Y^{\prime}-\frac{\mu}{\Lambda}Y\right)\ell_L \; , \\
E_R&=&E_{1R}^-
-\left(\frac{\mu-\mu^\prime}{4\Lambda}\right)E_{2L}^{+c}\; , \\
E'_L&=&E_{1L}^-
+\left(\frac{\mu-\mu^\prime}{4\Lambda}\right)E_{2R}^{+c}
-\frac{v}{\Lambda}Y \ell_L  \; , \\
E'_R&=&E_{2L}^{+c}
+\left(\frac{\mu-\mu^\prime}{4\Lambda}\right)E_{1R}^- \; .
\end{eqnarray}
where we have used that, in general, one can choose the flavour basis
such as $ V^\ell_L= V^\ell_R=I$.

To first order in the small parameters the neutral weak interactions 
of the
light states take the same form as that on the SM and the charged current
interactions read \footnote{Violation of unitarity (and flavour
  mixing) appears in the CC (and NC) interactions of the light leptons
  to higher order \cite{valle,gavelanu,Abada:2007ux}. }
\begin{eqnarray}
{\cal L}_W^{light}= -\frac{g}{\sqrt{2}}\left(
\overline{\ell_{L}}\gamma^\mu U_{LEP}\nu_{L}W_\mu^- \right) +h.c.
\end{eqnarray}
where $g$ is the $SU(2)_L$ coupling constant. 

After absorbing three
unphysical phases in the definition of the light charged leptons, the
leptonic mixing matrix can be chosen
\begin{eqnarray}
\!\!\!\!\!\!\!\!\!\!\!\!\!\!\!\!
U_{LEP}=V^\nu&=&
    \begin{pmatrix}
	1 & 0 & 0 \\
	0 & c_{23}  & {s_{23}} \\
	0 & -s_{23} & {c_{23}}
    \end{pmatrix}
\!\!\!
    \begin{pmatrix}
	c_{13} & 0 & s_{13} e^{-i\delta_\text{CP}} \\
	0 & 1 & 0 \\
	-s_{13} e^{i\delta_\text{CP}} & 0 & c_{13}
    \end{pmatrix}
\!\!\! 
    \begin{pmatrix}
	c_{21} & s_{12} & 0 \\
	-s_{12} & c_{12} & 0 \\
	0 & 0 & 1
    \end{pmatrix}
\!\!\!
        \begin{pmatrix}
	e^{-i \alpha} & 0 & 0 \\
	0 & e^{i \alpha} & 0 \\
	0 & 0 & 1
    \end{pmatrix},
\end{eqnarray}
where $c_{ij} \equiv \cos\theta_{ij}$ and $s_{ij} \equiv
\sin\theta_{ij}$.  The angles $\theta_{ij}$ can be taken without loss
of generality to lie in the first quadrant, $\theta_{ij} \in
[0,\pi/2]$ and the phases $\delta_\text{CP},\; \alpha\in [0,2\pi]$.
The leptonic mixing matrix contains only two phases because there are
only two heavy triplets and consequently only two light neutrinos are
massive while the lightest one remains massless.
\FIGURE{
\includegraphics[width=0.76\textwidth]{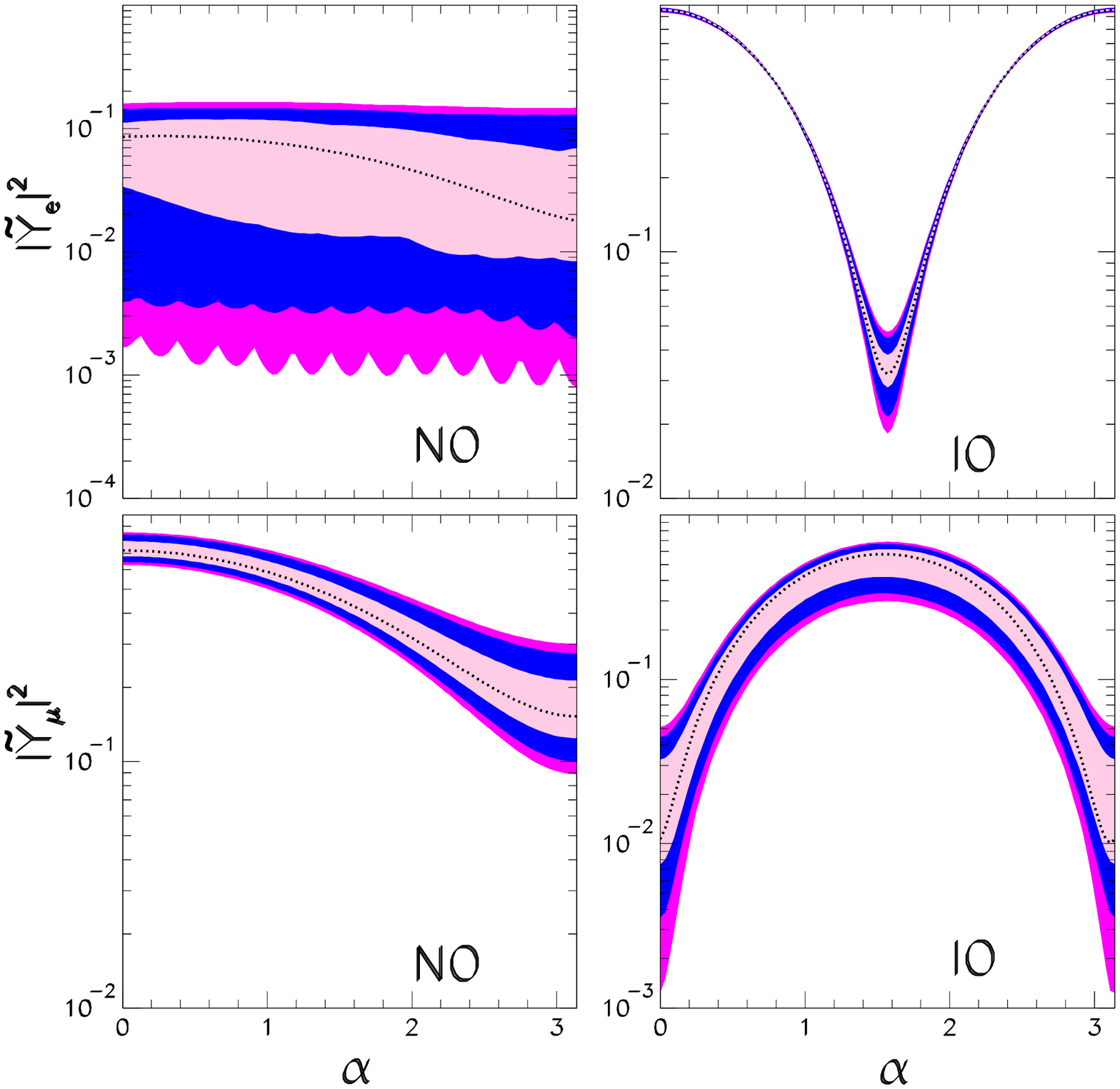}
\includegraphics[width=0.76\textwidth]{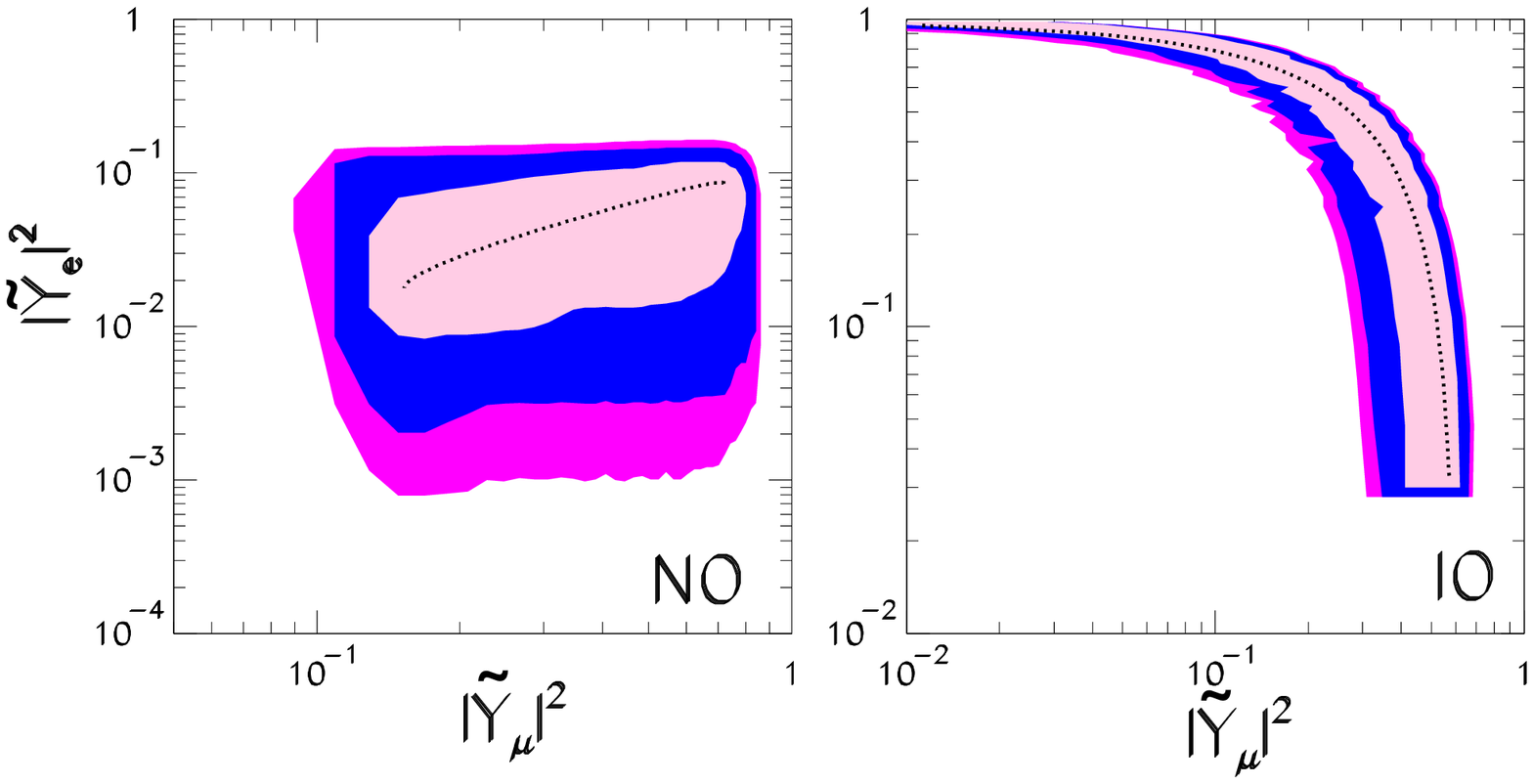}
\caption{ Allowed ranges of the Yukawa couplings $|\tilde Y_e|^2\equiv
  |Y_1|^2/y^2$ and $|\tilde Y_\mu|^2\equiv |Y_2|^2/y^2$ obtained from
  the global analysis of neutrino data \cite{ourfit}. The upper four
  panel shows the values of the couplings as a function of the unknown
  Majorana phase $\alpha$. The correlation between the two couplings
  is own in the two lower panels.  The left (right) panels correspond
  to normal (inverted) ordering. The dotted line corresponds to the
  best fit values. The ranges in the filled areas are shown at
  1$\sigma$, $2\sigma$, and 99\% CL.
  \label{fig:yuk}}}

As shown in Ref.~\cite{Gavela:2009cd} in this simple model one can
fully reconstruct the neutrino Yukawa coupling $Y$ and the combination
$\widehat {Y'}$ from the neutrino mass matrix. 
Therefore it is not possible to reconstruct the Yukawa couplings
$Y^\prime$ from Eq.~(\ref{eq:haty}) without
the knowledge of the parameters $\epsilon$ and $\mu$.
The reconstruction is different for normal and inverted orderings:
\begin{itemize}
\item Normal Ordering (NO):  In this case we have $0=m_1<m_2<m_3$
and the Yukawa couplings are given by
\begin{eqnarray}
Y_{a}=&&\frac{y}{\sqrt{2}}\left(\sqrt{1+\rho}\ U_{a3}^*+\sqrt{1-\rho}\ 
U_{a2}^*\right) \;,
\label{eq:no1}\\
\widehat {Y^\prime}_{a}=&&\frac
{\widehat{y^\prime}}{\sqrt{2}}\left(\sqrt{1+\rho}\ U_{a3}^*-
\sqrt{1-\rho}\ U_{a2}^*\right) \; , 
\nonumber
\end{eqnarray}
where $y$ and $\widehat {y'}$ are two real numbers and 
\begin{eqnarray}
&&\rho=\frac{\sqrt{1+r}-\sqrt{r}}{\sqrt{1+r}+\sqrt{r}}\; ,  
\;\;\;\;\;
r=\frac{m_2^2-m_1^2}{m_3^2-m_2^2}\; ,  
\label{eq:no2}
\\
&&m_1=0\;, \;\;\;\;\; \;\;\;\;\; \;\;\;\;\; 
m_2=\frac{\epsilon y {\widehat {y'}} v^2}{\Lambda}(1-\rho)\;, \;\;\;\;\;
m_3=\frac{\epsilon y {\widehat {y'}} v^2} {\Lambda}(1+\rho)\;.
\label{eq:no3}
\end{eqnarray}
\item Inverted Ordering (IO): If we have $0=m_3<m_1<m_2$ the Yukawa can 
be written as
\begin{eqnarray}
Y_{a}=&&\frac{y}{\sqrt{2}}\left(\sqrt{1+\rho}\ U_{a2}^*+\sqrt{1-\rho}\ 
U_{a1}^*\right)\;, 
\label{eq:io1}\\
\widehat {Y^\prime}_{a}=&&\frac{\widehat {y^\prime}}{\sqrt{2}}\left(\sqrt{1+\rho}\ U_{a2}^*-
\sqrt{1-\rho}\ U_{a1}^*\right)\; , 
\nonumber
\end{eqnarray}
with 
\begin{eqnarray}
&&\rho=\frac{\sqrt{1+r}-1}{\sqrt{1+r}+1}\; , \;\;\;\;\; 
r=\frac{m_2^2-m_1^2}{m_1^2-m_3^2}\; ,  
\label{eq:io2}
\\
&&m_3=0\;,  \;\;\;\;\;\;\;\;\;\; \;\;\;\;\;  
m_1=\frac{\epsilon y \widehat {y'} v^2}{\Lambda}(1-\rho)\;,
\;\;\;\;\; 
m_2=\frac{\epsilon y \widehat {y'} v^2} {\Lambda}(1+\rho)\;.
\label{eq:io3}
\end{eqnarray}
\end{itemize} 

We plot in Fig.~\ref{fig:yuk} the ranges of the Yukawa couplings
$|\tilde Y_e|^2\equiv |Y_1|^2/y^2$ and $|\tilde Y_\mu|^2\equiv
|Y_2|^2/y^2$ obtained by projecting the allowed ranges of oscillation
parameters from the global analysis of neutrino data \cite{ourfit}
using Eqs.~\eqref{eq:no1},~\eqref{eq:no2},~\eqref{eq:io1},
and~\eqref{eq:io2}.  The ranges are shown at 1$\sigma$, $2\sigma$, and
99\% CL (1dof) while the dotted line corresponds to the best fit
values. We show the ranges of these Yukawa couplings as a function of
the unknown Majorana phase $\alpha$, as well as we present the
correlation between the Yukawa couplings in the above flavours. As we
can see from this figure, the electron and muon Yukawa couplings
exhibit a quite different behavior with $\alpha$ for the NO and IO
cases. It is also interesting to notice that the two Yukawas are
invariant under $\alpha$ going into $\pi - \alpha$ in the limit that
$s_{13}$ or $\delta$ go to zero for the IO mass ordering.

Also to the same order the Lagrangian for the interactions of the heavy
triplet states reads:
\begin{eqnarray}
{\mathcal L}_W=&&-g\left(\overline{E^-_1}\gamma^\mu N W_\mu^- -
\overline{N}\gamma^\mu E^+_2W_\mu^-\right)+h.c. \nonumber \\
&&-\frac{g}{\sqrt{2}}\left(K_{a}
\overline{\ell_{aL}}\gamma^\mu N_LW_\mu^-
+\ K_a'\overline{\ell_{aL}}\gamma^\mu {N_{R}}^c
W_\mu^-\right ) +h.c.\nonumber\\
&&+g\left(
\tilde K_{a}\overline{\nu_{aL}} \gamma^\mu E^+_{2L} W_\mu^-
+\tilde K'_{a}\overline{\nu_{aL}} \gamma^\mu {E^-_{1R}}^c W_\mu^-\right) +h.c.
\label{eq:lw}
\\
{\mathcal L}_Z=&&gC_W
\left(
\overline{E^-_1}\gamma^\mu E^-_1Z_\mu-
\overline{E^+_2}\gamma^\mu E^+_2Z_\mu\right) \nonumber \\
&&+\frac{g}{2C_W}\left(\tilde K_{a}\overline{\nu_{aL}}\gamma^\mu N_{L} Z_\mu
+\tilde K'_{a}\overline{\nu_{aL}}\gamma^\mu {N_{R}}^c Z_\mu\right) +h.c. \nonumber
  \\
&&+\frac{g}{\sqrt{2}C_W}
\left( K_{a}\overline{\ell_{aL}}\gamma^\mu E^-_{1L}Z_\mu
+ K'_{a} \overline{\ell_{aL}}\gamma^\mu {E^+_{2R}}^cZ_\mu\right) +h.c.
\label{eq:lz}\\
{\mathcal L}_\gamma=&& e
\left(
\overline{E^-_1}\gamma^\mu E^-_1A_\mu-
\overline{E^+_2}\gamma^\mu E^+_2A_\mu\right) \\
{\mathcal L}_{h^0}=&&\frac{g\Lambda}{2M_W}\left(\tilde K_{a}
\overline{\nu_{aL}} N_{R}
+\tilde K^{\prime\prime}_a \overline{\nu_{aL}} {N_{L}}^c \right)h^0 +h.c. \nonumber \\
&&+\frac{g\Lambda}{\sqrt{2}M_W}\left( K_{a}\overline{\ell_{aL}}E^-_{1R}
+K^{\prime\prime}_i \overline{\ell_{aL}}{E^+_{2L}}^c\right)h^0 +h.c. \; .
\label{eq:lh}
\end{eqnarray}
where $c_W$ is the cosine of the weak mixing angle and the matrices
$K^{(\prime)(\prime)}$ and $\tilde K^{(\prime)(\prime)}$ 
are defined as:
\begin{equation}
\begin{array}{ll}
  K_a
  =-\frac{v}{\sqrt{2}\Lambda}  {Y_a}^{*} \;\;\; ,&  
  \tilde K_{a}={U^*_{LEP}}_{ca}  K_{c} \; ,
  \\
  K^{\prime}_a
  =-\frac{v}{\sqrt{2}\Lambda} \left[\epsilon {Y^{\prime}_a}^{*}-
   \left(\frac{3\mu+\mu^\prime}{4\Lambda}\right)Y_a^*\right] \;\;\; ,&  
  \tilde K^{\prime}_{a}={U^*_{LEP}}_{ca}  K^{\prime}_{c} \; .
  \\
  K^{\prime\prime}_a
  =-\frac{v}{\sqrt{2}\Lambda} \left[\epsilon {Y^{\prime}_a}^{*}-
   \left(\frac{\mu-\mu^\prime}{4\Lambda}\right)Y_a^*\right] \;\;\; ,&  
  \tilde K^{\prime\prime}_{a}={U^*_{LEP}}_{ca}  K^{\prime\prime}_{c} \; .
\end{array}
\label{eq:kdef}
\end{equation}

We can see from Eqs.~\eqref{eq:no1}--\eqref{eq:io3} that the flavour
structure of the lepton number conserving couplings of the heavy
triplet fermions, $K$ and $\tilde K$, is fully determined by the low
energy neutrino parameters.  Moreover, its strength is controlled by
the real number $y v/\Lambda$ while the combination $\epsilon y
\widehat {y'}/\Lambda$ is fixed by the neutrino masses.  On the other
hand we find that the $L$-violating couplings
$K^{\prime(\prime\prime)}$ and $\tilde K^{\prime(\prime\prime)}$
are different from the combination determined by 
the low energy neutrino parameters, $\widehat {Y'}$. This is, 
the  $L$-violating couplings of the triplet fermions are not fixed
by the low energy neutrino parameters. 
However one must notice these $L$-violating couplings are very
suppressed since in the MLFV framework the smallness of the neutrino
mass naturally stems from the smallness of total lepton number
violation which is associated with the smallness of the $\epsilon,
\mu$ and $\mu'$ parameters.

The low energy Lagrangian after
integrating out the triplet states takes the form of
Eq.~(\ref{eq:efflag}) with $\Lambda_{FL}=\Lambda$ and
$\Lambda_{LN}=\Lambda/\sqrt{\epsilon},\Lambda^2/\mu,\Lambda^2/\mu'$.
So there is no state with mass $\Lambda_{LN}$.  Furthermore the
hierarchy of scales $\Lambda_{LN}\gg \Lambda_{FL}$ is technically
natural in the t'Hooft's sense since it is associated with the
smallness of $\epsilon$, $\mu$ and $\mu'$ parameters and in the limit
$\mu,\mu',\epsilon\rightarrow 0$ total lepton number symmetry is
restored.
 
In what respects to the phenomenology of the heavy fermion triplets,
total lepton number violation appears in their decays as a consequence
of the presence of both ``primed'' and ``not primed'' couplings in
Eqs.~\eqref{eq:lw}--\eqref{eq:lh} as well as of the ${\cal
  O}(\mu/\Lambda,\mu'/\Lambda)$ mass splitting and mixing in the heavy
states. Thus small total lepton number violation implies a strong
hierarchy between the lepton number conserving and lepton number
non-conserving effects in the heavy fermion collider
phenomenology and renders  the observation of $L$-violating signals 
impossible at the LHC.
This is the main difference with the expected LHC
signatures in non MLFV scenarios for type-III see--saw such as the
ones studied for example in Refs.\cite{strumia, bajc} where $\Delta
L=2$ final states constitute a smoking gun signature which is very
suppressed in the MLFV model here considered. Consequently, when
discussing the signatures associated with this scenario, we will
concentrate on total lepton number conserving signals.


\section{Signatures}
\label{sec:sig}

The dominant production processes for the heavy triplet fermions in
this model are pair production due to their gauge interactions:
\begin{eqnarray}
pp\rightarrow E^+_i E^-_i ,&  \;\;\; & pp\rightarrow E^\pm_i N 
\;\;\; {\rm for}\; i=1,2\; ,
\end{eqnarray}
where for simplicity in the second reaction and in what follows we 
generically denote by  ``$N$'' either the $N$ or $\bar N$ state.   
The cross sections for these processes are well-known functions of
their mass, see for example \cite{delAguila:2008hw}, and for
completeness we plot them in the left panel of Fig.~\ref{fig:signals}.

\FIGURE[!t]{
\includegraphics[width=0.45\textwidth]{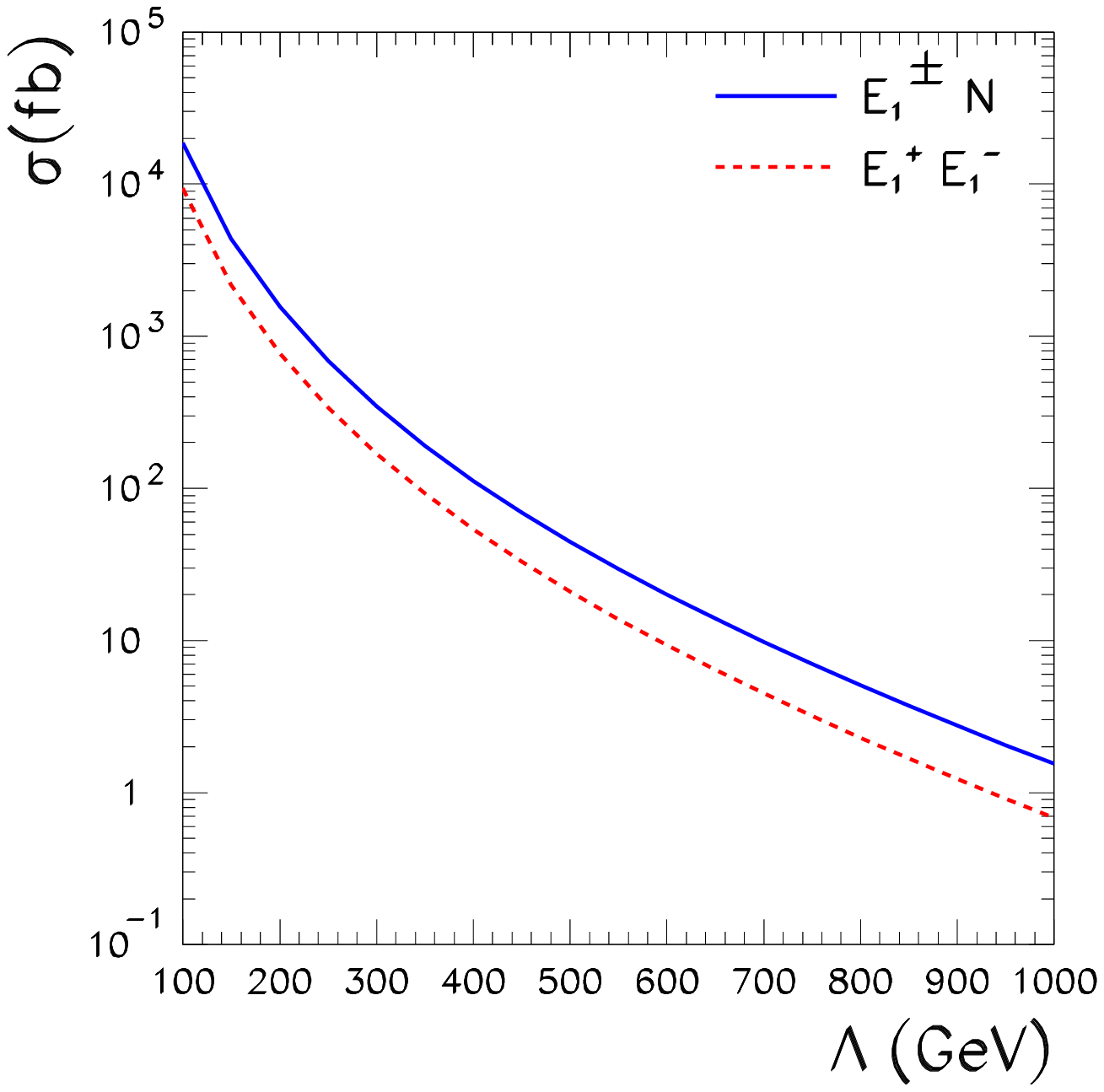}
\includegraphics[width=0.45\textwidth]{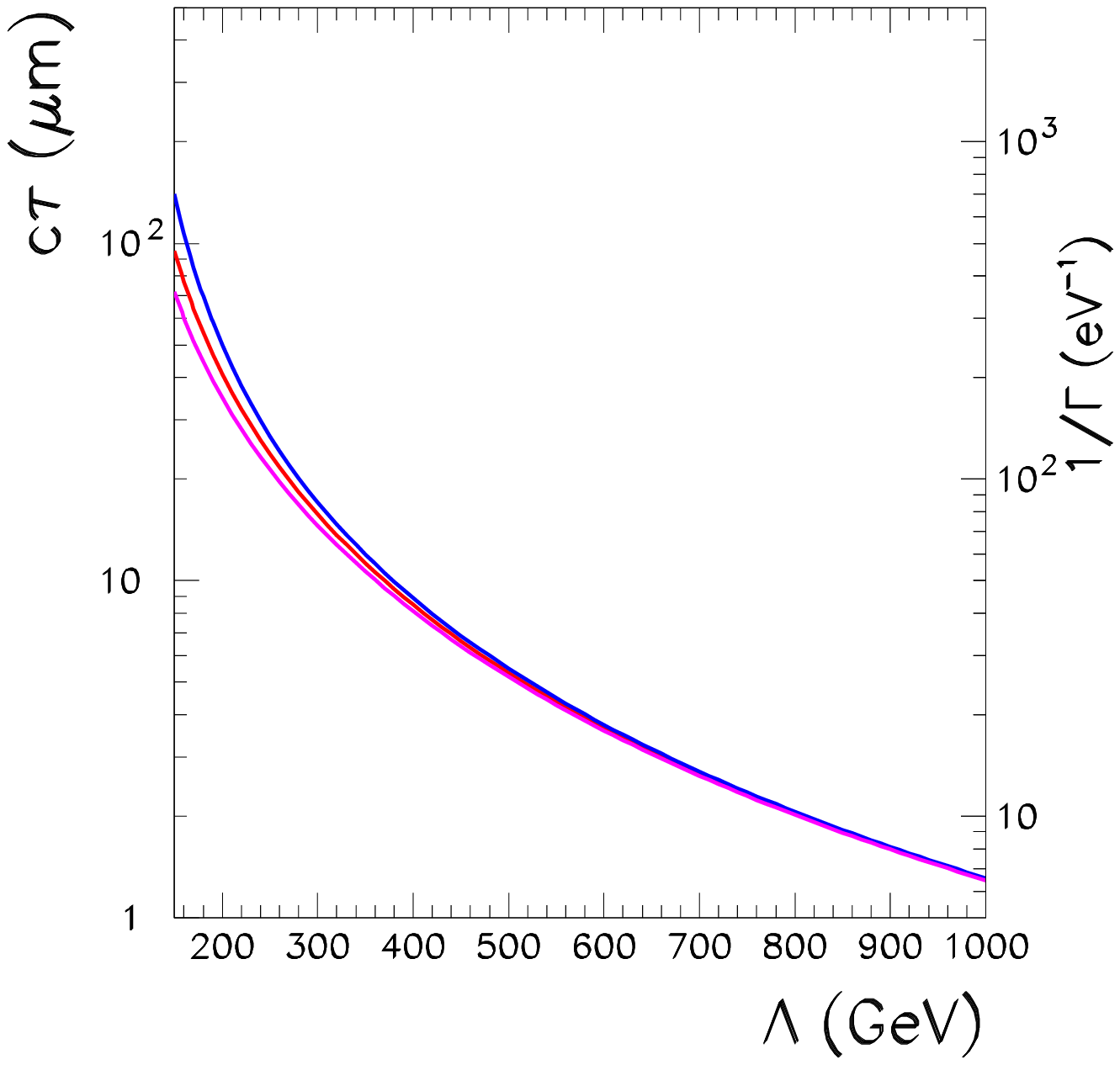}
\caption{{\bf Left panel:} Cross section for the triplet fermion pair
  production $N E_1^\pm$ and $E_1^+ E_1^-$ that have the same
  values that the cross sections for $N E_2^\pm$ and $E_2^+
  E_2^-$ respectively.  {\bf Right panel:} Maximum decay length of the
  triplet fermions $E^\pm_1$ (blue upper curve), $N$ (red middle
  curve), and $E^\pm_2$ (magenta lower curve). In all cases we have
  taken $m_{h^0}=120$ GeV and assumed $k=1/10$.}
\label{fig:signals}}

The widths for the different decay modes read~\cite{delAguila:2008hw}:
\begin{eqnarray}
&&\Gamma\left(N\rightarrow\ell_a^-W^+\right)=\frac{g^2}{64\pi}
|K_a|^2\frac{\Lambda^3}{M_W^2}
\left(1-\frac{M_W^2}{\Lambda^2}\right)\left(1+\frac{M_W^2}
{\Lambda^2}-2\frac{M_W^4}{\Lambda^4}\right) \; ,
\label{widths}\\
&&\Gamma\left(N\rightarrow\nu_a Z\right)=\frac{g^2}{128\pi
c_w^2}|\tilde K_a|^2\frac{\Lambda^3}{M_Z^2}
\left(1-\frac{M_Z^2}{\Lambda^2}\right)\left(1+\frac{M_Z^2}{\Lambda^2}
-2\frac{M_Z^4}{\Lambda^4}\right) \; ,
\nonumber\\
&&\Gamma\left(N\rightarrow\nu_a
h^0\right)=\frac{g^2}{128\pi}|\tilde K_a|^2\frac{\Lambda^3}{M_W^2}
\left(1-\frac{M_{h^0}^2}{\Lambda^2}\right)^2 \; ,
\nonumber\\
&&\Gamma\left(E_2^+\rightarrow\nu_a
W^+\right)=\frac{g^2}{32\pi}|\tilde K_a|^2\frac{\Lambda^3}{M_W^2}
\left(1-\frac{M_W^2}{\Lambda^2}\right)\left(1+\frac{M_W^2}{\Lambda^2}
-2\frac{M_W^4}{\Lambda^4}\right) \; ,
\nonumber\\
&&\Gamma\left(E_1^-\rightarrow\ell_a^-Z\right)=\frac{g^2}{64\pi
c_W^2}|K_a|^2\frac{\Lambda^3}{M_Z^2}
\left(1-\frac{M_Z^2}{\Lambda^2}\right)\left(1+\frac{M_Z^2}{\Lambda^2}
-2\frac{M_Z^4}{\Lambda^4}\right) \; ,
\nonumber\\
&&\Gamma\left(E_1^-\rightarrow\ell_a^-h^0\right)=\frac{g^2}{64\pi}
| K_a|^2\frac{\Lambda^3}{M_W^2}
\left(1-\frac{M_{h^0}^2}{\Lambda^2}\right)^2 \; .
\nonumber
\end{eqnarray}
Using Eq.~\eqref{eq:kdef} it is trivial to show that
\begin{eqnarray}
\sum_{a=1}^3|K_a|^2=\sum_{a=1}^3|\tilde K_a|^2=\frac{y^2 v^2}{2\Lambda^2} 
\; ,
\end{eqnarray}
and show that total decay widths for the three triplet fermions
$F=N,E^-_1,E^+_2$ are
 \begin{eqnarray}
&&\Gamma_{F}^{\mbox{\begin{tiny}TOT\end{tiny}}}
=\frac{g^2 \Lambda^3}{64 \pi M_W^2} \frac{y^2 v^2}{\Lambda^2} 
(1+{\cal F}_F(\Lambda))  
\label{totalwidths}
\end{eqnarray}
where ${\cal F}_{F}(\Lambda)\rightarrow 0$ for $\Lambda\gg
m_{h^0},M_Z,M_W$. In a general Type-III see--saw model it is possible
that the branching ratio of $N$ or $E^\pm_i$ into a light lepton of a
given flavour is vanishingly small. This is not the case for the
Type-III see--saw MLFV model studied here since the Yukawa couplings
fixed by the neutrino physics are non-vanishing; see
Fig.~\ref{fig:yuk}.

Other important characteristic of this simple MLFV model is that the 
values of the neutrino masses imply a lower bound on the total decay 
width of the triplet fermions as a consequence of the hierarchy between
the $L$-conserving and $L$-violating $y$ and $\epsilon \widehat{y'}$ 
constants. Let us write $\epsilon \widehat {y'} < k y$, where $k<1$. From
Eq.~\eqref{eq:no3} or \eqref{eq:io3} it follows that
\begin{eqnarray}
\frac{y^2 v^2}{\Lambda^2}>\frac{ m_{3(2)}}{k \Lambda(1+\rho)}=
\frac{ \sqrt{m_{3(2)}^2-m_{1(3)}^2}}{k \Lambda (1+\rho)}> 
\frac{0.046\ {\rm eV}}{k \Lambda}
\label{eq:minw}
\end{eqnarray}
where the last number is obtained at 99\% CL from the global analysis
of neutrino data~\cite{ourfit}. We depict in the right panel of
Fig.~\ref{fig:signals} the resulting minimum decay width for the
triplet fermions as well as their corresponding maximum decay length 
for any value of $k<0.1$.
From this figure we see that in this minimal model,
even for heavy states as light as $\Lambda=150$ GeV, the corresponding
decay length is always
\begin{equation}
c \tau \lesssim 100 \,{\rm \mu m}
\end{equation}
and it decreases rapidly with $\Lambda$. Such a small decay length is
too short to produce a detectable displaced decay
vertex~\cite{Aad:2008zzm,:2008zzk}. The use of detached vertices as
signatures of the heavy state decays have been discussed in the
context of more general see--saw
models~\cite{han,strumia,li,bajc}. The lack of this signature in this
MLFV model makes the background reduction more challenging.
Conversely if a triplet fermion signal is found without a displaced
vertex, it will point out towards a very hierarchical neutrino
spectrum such as predicted in this simple model.

The most characteristic signature of MLFV Type-III see--saw models is
the dependence of the decays of the triplet fermions on the low energy
neutrino parameters through the Yukawa couplings.  Therefore, in order
to be able to tag the lepton flavours, we are lead to consider
processes where the new fermions have two-body decays exhibiting
charged leptons, {\em i.e.}
\begin{eqnarray}
pp\rightarrow F (\rightarrow \ell_a X) F' (\rightarrow \ell_b X')  
\end{eqnarray}
for $F$,$F'=N,E_i$ and with $X,X'=Z,W,h^0$. In fact, it turns out
that the production cross sections of these processes satisfy
\begin{equation}
\sigma \left[pp\rightarrow F (\rightarrow l_a X) 
F' (\rightarrow l_b X')\right] 
\propto |\tilde{Y}_a|^2|\tilde{Y}_b|^2 
\label{eq:sig0}
\end{equation}
where $\tilde{Y}_a\equiv\frac{Y_a}{y}$.  Therefore, the number of
events for final states with different combinations of charged lepton
flavours ($a$, $b$) can be fully determined in terms of the low-energy
neutrino parameters. However, in order to test this prediction one
must take into account how the SM bosons, $X$ and $X'$, decay and that
the final state contains at least six particles, what makes the
reconstruction of the decay chain non trivial, as well as, the
presence of the irreducible SM backgrounds.

Keeping in mind the above discussion, the most promising signatures to
both detect the triplet fermions, as well as, to test the flavour
predictions in this model are those in which
\begin{itemize}

\item[(i)] The branching ratios into the final state after considering
  the decays of $X, X'$ are not strongly suppressed.

\item[(ii)] After reconstruction the process should allow us to
  identify the charged leptons $\ell_{i,j}$ originating from the two
  body decays of the triplet fermions.

\item[(iii)] To have further information the topology should permit
  the identification of the bosons $X$ or $X'$.

\item[(iv)] We should be able to reconstruct the invariant mass of the
  systems $X l$ to identify the presence of the triplet fermion pair.


\end{itemize}

Altogether we find that the most promising final states,
that can be fully reconstructed, are three leptons plus two jets and
missing energy proceeding via
\begin{eqnarray}
&&pp\rightarrow W^\pm\rightarrow 
N (\rightarrow l_a^\mp W^\pm\rightarrow l_a^\mp l_c^\pm \nu)
\ E_1^\pm(\rightarrow\ell_b^\pm Z/h \rightarrow \ell_b^\pm j  j) \; , 
\label{eq:proc1}
\end{eqnarray}
and two leptons and four jets resulting from
\begin{equation}
\begin{array}{l}
pp\rightarrow W^\pm\rightarrow 
N (\rightarrow \ell_a^\mp W^\pm\rightarrow \ell_a^\mp j j )  \;
\ E_1^\pm (\rightarrow \ell_b^\pm Z/h^0\rightarrow \ell^\pm_b jj) \; ,
\\
pp\rightarrow Z/\gamma\rightarrow E_1^\mp 
(\rightarrow\ell_a^\mp Z/h^0\rightarrow \ell_a^\mp jj)  
E_1^\pm(\rightarrow\ell_b^\pm Z/h^0\rightarrow \ell_b^\pm jj) \; .
\end{array}
\label{eq:proc2}
\end{equation}
In order to establish the observability of these signals it is
important to keep in mind that the final states present not only SM
backgrounds, but also receive contributions from other decays of the
triplet fermions, as we shall see. Notice also that we do not consider
the production of the charged heavy fermions $E_2^\pm$ since they
decay exclusively into $\nu W$ pairs, so flavour tagging of the final
leptons it is not possible. Such processes can contribute to extend
the LHC potential to unravel the existence of the triplet fermions but
do not allow for the test of the MLFV hypothesis.

We study process \eqref{eq:proc1} in detail in Sec.~\ref{sec:llljjn}.
It is characterized by a good signal to background
ratio~\cite{delAguila:2008cj} and the main challenges, as we will see,
are the reconstruction conditions $(ii)$ and $(iv)$. Process
\eqref{eq:proc2} is analyzed in Sec.~\ref{sec:lljjjj}. In this case
both bosons decay hadronically what gives a high signal rate and since
there are only two leptons in the final state and no neutrinos, the
reconstruction conditions are more easily fulfilled. The main
challenge, as we will see is the presence of a larger QCD background.

We perform our analysis at the parton level, keeping the full helicity
structure of the amplitude for both signal and backgrounds. This is
achieved using the package MADGRAPH~\cite{madevent} modified to
include the new fermions and their couplings. In our calculations we
use CTEQ6L parton distribution functions \cite{CTEQ6} and the MADEVENT
default renormalization and factorization scales and a $pp$ center of
mass energy $\sqrt{s}=14$ TeV, unless otherwise stated. Furthermore,
we simulate experimental resolutions by smearing the energies, but not
directions, of all final state leptons and jets with a Gaussian error
given by a resolution $\Delta E/E=0.14/\sqrt{E}$ for leptons while for
jets we assumed a resolution $\Delta E/E=0.5/\sqrt{E}\oplus0.03$, if
$|\eta_j|\leq 3$ and $\Delta E/E=1/\sqrt{E}\oplus0.07$, if
$|\eta_j|>3$ ($E$ in GeV).  We also consider a lepton detection
efficiency of $e^\ell=0.9$ and a jet one of $e^j=0.75$. For
simplicity, we assumed the Higgs mass to be 120 GeV in all our
analyses.

\section{Process $pp\rightarrow\ell\ell\ell jj \Sla{E_T}$}
\label{sec:llljjn}

In this section we study the process  
\begin{equation}
pp \rightarrow \ell_1^\mp \,\ell_2^\pm \,\ell_3^\pm\, j \, j \, \Sla{E_T}
\label{chan1}
\end{equation}
where we focus on final leptons being either electrons or muons for
easier flavour tagging.  This final state  allow
us to look for the events originating from production of triplet
fermions in  Type-III see--saw models as shown in
Eq.~\eqref{eq:proc1}.

The dominant irreducible SM backgrounds are:
\begin{enumerate}

\item $t\bar t W$ production where the two b from $t \rightarrow W b$
  decay are identified as the jets and the three $W's$ decay
  leptonically; 

\item $t\bar t Z$ where the $Z$ decays leptonically while one top
  decays semi-leptonically and the other decays fully hadronically.
  Another possibility is that the two top quarks decay
  semi-leptonically, however, one of the four final state leptons is
  lost or misidentified.  This background can contain up to 4 jets, in
  addition to the three leptons, of which we require that at least two
  comply with the acceptance cuts described below; see
  Eq.~\eqref{eq:basiccuts};

\item $WZ jj$ and $ZZjj$ with both $W$ and $Z$ decaying leptonically
  and one lepton in the $ZZjj$ case escapes detection.

\end{enumerate}
In principle the backgrounds from channels containing leptonic $Z$
decays can be reduced by vetoing events where the opposite-sign
equal-flavour leptons have invariant mass close to the $Z$ mass~
\cite{delAguila:2008cj,nir}. However, as we will see, in this MLFV
model signals are large only for relatively light triplet fermions,
$\Lambda \leq 500$ GeV, and for these masses the characteristic
invariant mass of the opposite-sign equal-flavour lepton pair is not
far from the $Z$ mass either. So the $Z$ veto reduces also the signal
and no gain in the observability is obtained; for an illustration see
Fig.~\ref{fig:p1:mll}.  Additional backgrounds, like $t\bar t$ and $Z
b\bar b$, with additional leptons produced from the semi-leptonic
decays of the $b's$ are negligible when no $Z$ veto is applied.
Furthermore, we did not take into account reducible backgrounds
stemming from the misidentification of a jet as a lepton.

\FIGURE[!t]{
\includegraphics[width=3.5in]{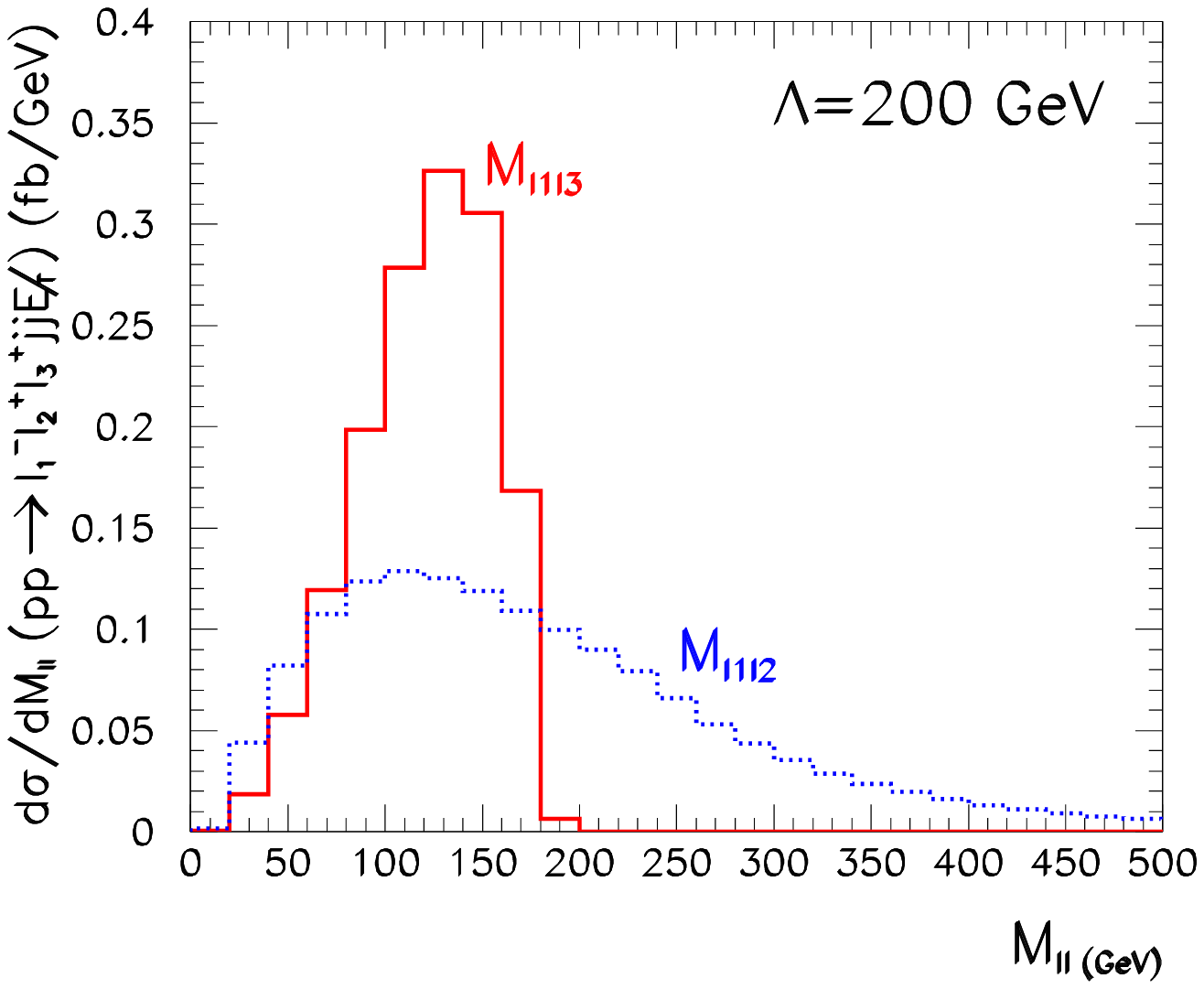}
\caption{Invariant mass distribution of the two possible
  opposite-sign equal-flavour lepton pairs for the signal
  $pp\rightarrow\ell\ell\ell jj \Sla{E_T}$.  Here we considered a
  heavy fermion mass $\Lambda=200$ GeV.}
\label{fig:p1:mll}}

We start our analysis by applying the following acceptance cuts, that
are meant to ensure the detection and isolation of the final leptons
and jets, as well as a minimum transverse momentum
\begin{eqnarray}
|\eta_\ell| < 2.5 \;\;,\;\; |\eta_j| < 3 \;\;,\;\;\;&
\Delta R_{\ell \ell},\Delta R_{\ell j},\Delta R_{jj},
> 0.4 \;\; &,\;\;\;
p_{T}^\ell,\;p_{T}^j  > 20 \hbox{ GeV}  \:,
\label{eq:basiccuts}
\end{eqnarray}
and a minimum missing transverse energy 
\begin{equation}
\Sla{E}_T > 10 \hbox{ GeV}\; .
\end{equation}
Next, we look for the two jets to be compatible with a $Z$ or a $h^0$
{\em i.e.}
\begin{equation}
M_Z-10\;\hbox{ GeV} < M_{jj} < m_{h^0}+10\; \hbox{ GeV.}
\label{eq:mjjzh}
\end{equation}

Our reconstruction procedure aims to single out events that originate
from the reaction \eqref{eq:proc1} in order to test the MLFV hypothesis,
therefore it is not optimized to
get the full LHC potential for the heavy triplet fermion discovery.
In order to reconstruct the $E_1^\pm$ and $N$ states we need to
identify which of the equal sign leptons $\ell_{2,3}$, is produced in
the $E_1^\pm$ two-body decay, as well as, which lepton comes from the $W$ in the
$N$ decay chain. To do so we start by reconstructing the two
possible values of the invariant mass of each of the equal sign lepton
plus two jets, $M_{\ell_2 jj}$ and $M_{\ell_3 jj}$.  If both
$M_{\ell_2 jj}$ and $M_{\ell_3 jj}$ are incompatible with the heavy
fermion mass, {\em i.e.}
\begin{equation}
M_{\ell_2 jj}
, M_{\ell_3 jj} \in\!\!\!\!\!
\slash\;\;
\,
\left(\Lambda-40, \Lambda+40\right)\mbox{GeV}
\label{eq:minvcut}
\end{equation}
the event is discarded. If only one of the two reconstructions is
inside this range we consider the corresponding lepton as the one
coming from $E_1^\pm$.  If both $M_{\ell_2 jj}$ and $M_{\ell_3 jj}$
are inside the range given in Eq.~\eqref{eq:minvcut} we proceed to
reconstruct the momentum of the neutrino using that in this final
state the neutrino momentum can be reconstructed up to a two--fold
ambiguity: its transverse momentum can be directly obtained from
momentum conservation in the transverse directions while its
longitudinal component can be inferred by requiring that
$(p_\nu+p_{\ell_k})^2 = M_W^2$ that leads to
\begin{eqnarray}
\!\!\!\!\!\!\!\!\!\!
p_L^{\nu_{k,n}}&=&\frac{1}{2 {p^{\ell_k}}^2}
\bigg\{\big[M_W^2+2(\vec {p^{\ell_k}_T} \cdot \vec{\Sla{p_T}})\big] 
p_L^{\ell_k} 
\pm 
\sqrt{\big[M_W^2+2(\vec{p_T^{\ell_k}}
 \cdot \vec{\Sla{p_T}})\big]^2 |\vec p^\ell|^2 -
4 (p_T^{\ell_k} E^{\ell_k}\Sla{E_T})^2}\bigg\}\;
\label{eq:plnu}
\end{eqnarray}
for $k=2,3$ and we label $n=1,2$ the solutions with $+,-$
respectively.  If neither $\ell_2$ nor $\ell_3$ lead to a real value of
Eq.~\eqref{eq:plnu}, the event is rejected.  If only one of them has
an acceptable solution we classify this lepton as the one coming from
$W$.

Finally if both leptons lead to satisfactory solutions of
Eq.~\eqref{eq:plnu} we proceed to reconstruct the neutral heavy fermion
$N$.  For each $\ell_{2,3}$, and using the two possible solutions for
the momentum of the neutrino $p^{\nu_{k,n}}$ ($k=2,3$ $n=1,2$) we
evaluate four invariant masses $M_ {\ell_1\ell_k \nu_{k,n}}$.  If for
both $k=2$ and $k=3$ the two $M_ {\ell_1\ell_k \nu_{k,1}}$ and $M_
{\ell_1\ell_k \nu_{k,2}}$ are outside the interval $\left(\Lambda-40,
  \Lambda+40\right)\mbox{GeV}$ we do not consider the event. If only
$k=2$ or $k=3$ has at least one of the corresponding $M_ {\ell_1\ell_k
  \nu_{k,n}}$ inside this range we select $\ell_k$ as the one coming
from $W$. Finally, if the ambiguity is still there and both leptons
have at least one solution inside this range we cut out the event.
In the cases where we identify the leptons before using the reconstruction
of the invariant mass of $N$ we also require at the end at least one
of the two possible reconstructions to be inside the range
$(\Lambda-40, \Lambda+40)$ GeV.

We illustrate the efficiency of this reconstruction procedure in
Fig.~\ref{fig:mrec}. In the left panel we depict the invariant mass
$M_{\ell jj}$ distribution for the signal (empty back histogram) and
the background (filled blue histogram) where we averaged over the two
possible combinations with $\ell=\ell_2$ or $\ell=\ell_3$. In this
plot we imposed the cuts in
Eqs.~\eqref{eq:basiccuts}--\eqref{eq:mjjzh} before the reconstruction
of the $E_1^\pm$ and $N$ states. On the right panel we present
the reconstructed invariant mass of the selected combination after the
procedure described above. As seen in the figure the procedure selects
most of the right combination for the $E_1^\pm$ signal peak while
efficiently reducing the background.

\FIGURE[!t]{
\includegraphics[width=0.85\textwidth]{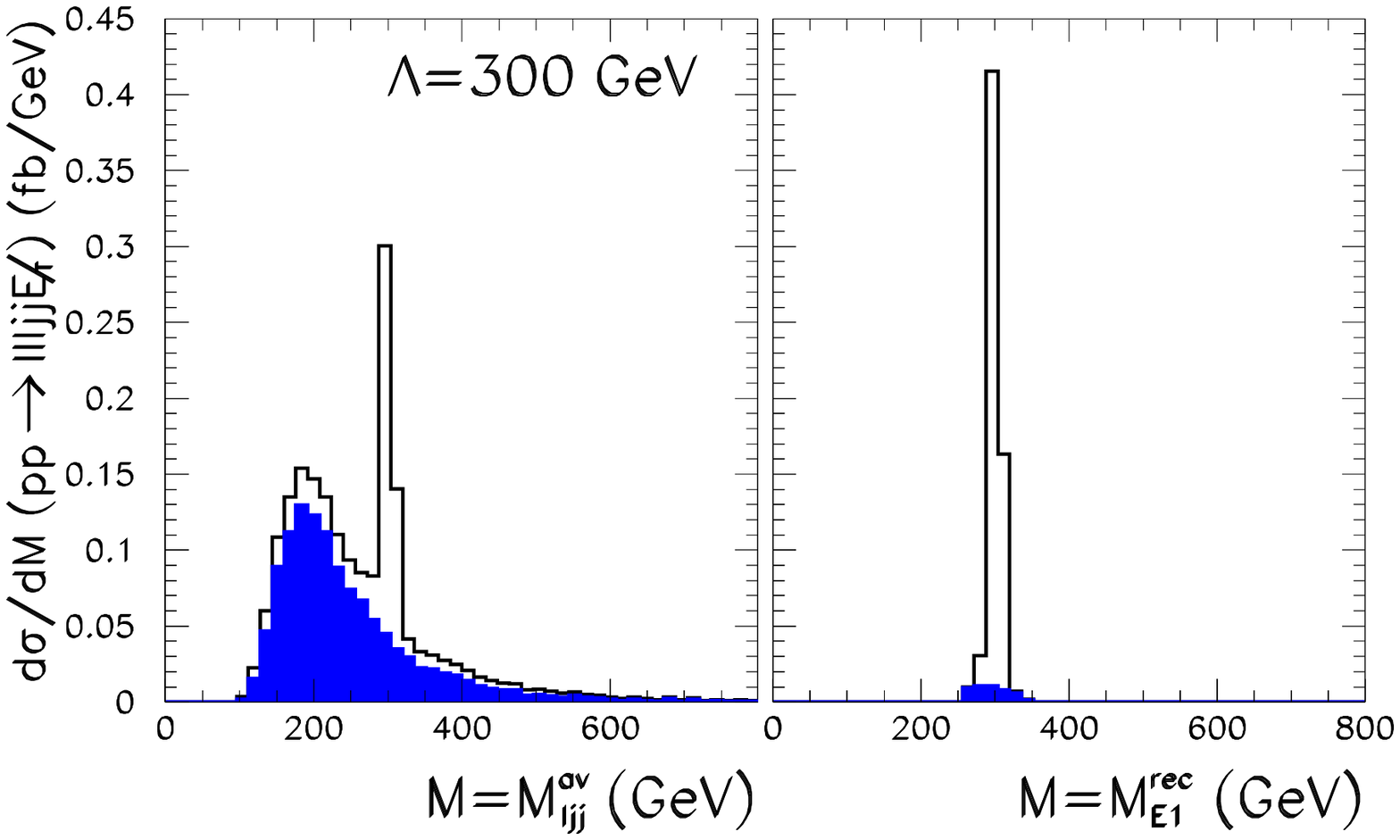}
\caption{Invariant mass distribution for signal (empty back histogram)
  and background (filled blue histogram) $M_{\ell jj}$. In the left
  panel we show the distribution averaged over the two possible
  combinations with $\ell=\ell_2$ or $\ell=\ell_3$ after imposing the
  cuts in Eqs.~\eqref{eq:basiccuts}--\eqref{eq:mjjzh} and before the
  reconstruction of the $E_1^\pm$ and $N$ states. On the right
  panel we show the $E_1^\pm$ reconstructed invariant mass after the
  selection procedure described in the text. The figure is shown for
  $\Lambda=300$ GeV and for characteristic values of the neutrino
  parameters: $\Delta m^2_{31}=2.4\times 10^{-3}$ eV$^2$ (NO), $\Delta
  m^2_{21}=7.65\times 10^{-5}$ eV$^2$, $\sin^2\theta_{23}=0.5$,
  $\sin^2\theta_{12}=0.304$ and $\sin^2\theta_{13}=0.03$ and vanishing
  values of the phases $\alpha=\delta_{\rm CP}=0$ (for these
  parameters, $\tilde Y_e=0.37$ and $\tilde Y_\mu=0.84$).}
\label{fig:mrec}}

In Ref.~\cite{nir} the ambiguity in the assignment of the equal-sign
leptons to the heavy lepton or the $W$ decays was resolved associating
to the $W$ decay the lepton that leads to the smallest transverse mass
\begin{equation}
M_T^W =
\sqrt{2p_T^{\ell_k}\Sla{E}_T\left(1-\cos\Phi_{\ell_k\Sla{E_T}}\right)} \; ,
\end{equation}
where $\Phi_{\ell_k\Sla{E_T}}$ is the angle between the lepton and the
missing energy.  We verified that this procedure is almost equivalent
to ours for high triplet masses $\Lambda$. Notwithstanding, for
lighter $\Lambda$ our reconstruction procedure selects more often the
correct lepton configuration and after applying the cuts on the
invariant masses of $N$ and $E^\pm_1$ it renders a better signal to
the background ratio.  For example for $\Lambda=200$ GeV our
reconstruction procedure leads to a misidentification probability of
2\% while using only the transverse invariant mass ordering this is
increased to 12\%.

After cuts and our reconstruction procedure the total cross section
of process \eqref{eq:proc1} can be written as
\[
    \sigma_0 (2 - \delta_{ab}) |\tilde{Y_a}|^2 |\tilde{Y_b}|^2
\]
when we generate events with the flavour combination $ab$. As we will
shortly see, most of these events are classified as having the correct flavour
combination $ab$ by our selection procedure, however, a fraction of
them are misidentified and labeled $ac$ for $b\ne c$ with a cross
section $\sigma_1$. 
This happens because we assign wrongly to the triplet fermion
a same-sign lepton with a different flavour coming from $W$.
Notice that both classes of events are exclusive since we reject
through the reconstruction procedure events that are compatible simultaneously
with the $ab$ and $ac$
flavour combinations. Furthermore, Eq.~\eqref{eq:proc1} is not the
only signal process leading to the final state of Eq.~\eqref{chan1} in the case we have
two opposite sign leptons of the same flavour. In this case there are also contributions from:
\begin{equation}
\begin{array}{l}
pp\rightarrow W^\pm\rightarrow 
N(\rightarrow \nu_m/\bar\nu_m Z\rightarrow \nu_m/\bar\nu_m 
\ell^+_a\ell^-_a)
\ E_1^\pm(\rightarrow\ell_b^\pm Z/h^0 \rightarrow \ell_b^\pm j  j) \; , 
\\
pp\rightarrow W^\pm\rightarrow 
N(\rightarrow \nu_m/\bar\nu_m Z/h^0\rightarrow \nu_m/\bar\nu_m j j)
\ E_1^\pm(\rightarrow\ell_b^\pm Z \rightarrow \ell_b^\pm \ell^+_a\ell^-_a
) \; .
\end{array}
\label{eq:otherproc}
\end{equation}
It is easy to see that summing over the undetectable neutrino type $m$
the cross section for these processes is proportional to
$|\tilde{Y}_b|^2$ and we denoted it by $\sigma_2
|\tilde{Y}_b|^2$. These events are classified as $aa$ flavour
combination with a cross section $\sigma_3 |\tilde{Y}_b|^2$, or as
$ab$ with a cross section $(\sigma_2-\sigma_3) |\tilde{Y}_b|^2$.  So
altogether the expected  signal ($S$) cross section in each
flavour channel is:
\begin{equation}
\begin{array}{l}
\sigma^{S}_{ee}= \left(\sigma_0-\sigma_1\right)|\tilde{Y}_e|^4
+\sigma_1|\tilde{Y}_e|^2|\tilde{Y}_\mu|^2+
\sigma_2 |\tilde{Y}_e|^2+\sigma_3 |\tilde{Y}_\mu|^2
\; ,
\\
\sigma^{S}_{\mu\mu}=
\left(\sigma_0-\sigma_1\right)|\tilde{Y}_\mu|^4+\sigma_1|\tilde{Y}_e|^2|
\tilde{Y}_\mu|^2+
\sigma_2|\tilde{Y}_\mu|^2+\sigma_3|\tilde{Y}_e|^2
\; ,
\\
\sigma^{S}_{e\mu}= \sigma_1\left(|\tilde{Y}_e|^4+|\tilde{Y}_\mu|^4\right)
+2\left(\sigma_0-\sigma_1\right)|\tilde{Y}_e|^2|\tilde{Y}_\mu|^2
+\left(\sigma_2-\sigma_3\right)\left(|\tilde{Y}_e|^2+|\tilde{Y}_\mu|^2\right)
\; ,
\\
\sigma^{S}_{TOT}= \sigma_0\left(|\tilde{Y}_e|^4+|\tilde{Y}_\mu|^4
+2  |\tilde{Y}_e|^2|\tilde{Y}_\mu|^2\right)
+2\sigma_2\left(|\tilde{Y}_e|^2+|\tilde{Y}_\mu|^2\right)
\; .
\end{array}
\label{eq:signall}
\end{equation}

\TABLE[!ht]{
\centering
\begin{tabular}{||c||c|c|c|c||}\hline
$\Lambda$(\ GeV) & $\sigma_0$ & $\sigma_1$ & $\sigma_2$ & $\sigma_3$ \\ \hline
150 & 80.2 & 2.05 & 7.53 & 1.78  \\ \hline
200 & 44.2 & 0.417 & 2.20 & 0.625 \\ \hline
300 & 12.9 & 0.027 & 0.125 & 0.043\\ \hline
500 & 1.90 & $<10^{-2}$ 
& $<10^{-2}$  & $<10^{-2}$  \\ \hline
\end{tabular}
\caption{Contributions to the signal cross sections ($\sigma^S_{ab}$)
  in fb for the processes $pp \rightarrow \ell^\mp \,\ell^\pm
  \,\ell^\pm\, j \, j \, \Sla{E_T}$ for $\ell=e,\mu$ according to
  Eq.~\eqref{eq:signall}. The results are presented for different
  values of the triplet fermion mass $\Lambda$ and they do not include
  the detection efficiencies for the leptons and jets.
\label{tab:sig1}}}

We present in Table~\ref{tab:sig1} the different contributions to the
signal cross section $\sigma^S_{ab}$ (in fb) after cuts
\eqref{eq:basiccuts}--\eqref{eq:mjjzh} and the triplet fermion
reconstruction for several values of $\Lambda$.  As we can see, the
bulk of the events passing our cuts originate from correctly
reconstructing process \eqref{eq:proc1}.  The SM background cross
sections $\sigma^B_{ab}$ are given in Table~\ref{tab:bck1}, where we
can see that the dominant SM background is $WZjj$ production.

\TABLE[!ht]{
\centering
 \begin{tabular}{||c|c|c|c|c|c|c||}\hline
& 
\multicolumn{2}{c|}  {$\Lambda=150$GeV}& 
\multicolumn{2}{c|}   {$\Lambda=200$GeV}& 
\multicolumn{2}{c||}  {$\Lambda=300$GeV} \\ \hline
 Process & 
$\sigma^B_{ee}=\sigma^B_{\mu\mu}$ & $\sigma^B_{e\mu}$&
$\sigma^B_{ee}=\sigma^B_{\mu\mu}$ & $\sigma^B_{e\mu}$&
$\sigma^B_{ee}=\sigma^B_{\mu\mu}$ & $\sigma^B_{e\mu}$\\ \hline
 $pp\rightarrow t\bar{t}W$ & 0.016 & 0.037
&0.021 & 0.045
&0.003  & 0.005
\\ \hline
 $pp\rightarrow t\bar{t}Z$ & 0.082 & 0.068
& 0.115 & 0.074 
& 0.036& 0.011
\\ \hline
 $pp\rightarrow WZjj$ & 1.66 & 1.15
& 1.58 & 0.950
& 0.27 & 0.118
\\ \hline
 $pp\rightarrow ZZjj$ & 0.04 & 0.022
& 0.046 & 0.028
  & 0.006 & 0.002
\\ \hline
  Total & 1.80   & 1.28  
&1.76 & 1.10 
&0.31 &  0.14
\\ \hline
 \end{tabular}
\caption{ SM background cross sections ($\sigma^B_{ab}$) in fb for the
  processes $pp \rightarrow \ell^\mp \,\ell^\pm \,\ell^\pm\, j \, j \,
  \Sla{E}_T$ with $\ell =e, \mu$.  The results are presented for
  different values of the triplet fermion mass $\Lambda$ and they do
  not include the detection efficiencies for the leptons and jets.}
\label{tab:bck1}}

We are now in position to evaluate the expected number of signal $(S)$ 
and background $(B)$ events for
the $\ell\ell\ell jj\Sla{E}_T$ topology with a given flavour
combination $ab$ as a function of the neutrino mass and mixing
parameters. This can be easily obtained from Eq.~\eqref{eq:signall},
Table \ref{tab:sig1}, and using the values of the Yukawa couplings in
\eqref{eq:no1}--\eqref{eq:io3}
\begin{equation}
N^{S,B} _{ab}=\sigma^{S,B}_{ab} \times {\cal L} \times \epsilon \; ,
\label{eq:nev}
\end{equation}
where ${\cal L}$ is the integrated luminosity and $\epsilon =
{e^l}^3\times {e^j}^2=0.41$ is the detection efficiency for leptons
and jets. 

Clearly the number of signal events depends on the value of the
triplet mass $\Lambda$ as well as on the neutrino parameters, which we
denote here by $\vec\theta=(\theta_{12},\theta_{13},\theta_{23},
\Delta m^2_{21},\Delta m^2_{31},\delta)$ and the Majorana phase
$\alpha$.  For example, we present in Fig.~\ref{fig:nev200} the range
of predicted number of events in the different flavour combinations
for a triplet fermion of mass $\Lambda=200$ GeV and an integrated
luminosity of ${\cal L}=30$ fb$^{-1}$, where the result is shown as a
function of the unknown Majorana phase $\alpha$ and the other neutrino
parameters $\vec\theta$ are obtained from the global analysis of
neutrino data~\cite{ourfit}. The left (right) panels correspond to
normal (inverted) ordering. The ranges are shown for values of
$\vec\theta$ allowed at 1$\sigma$, $2\sigma$, and 99\% CL while the
dotted line corresponds to the best fit values.  The horizontal dashed
lines are the corresponding number of SM background events as obtained
from Eq.~\eqref{eq:nev} with cross sections $\sigma^B_{ab}$ in Table
\ref{tab:bck1}.

It is important to notice from Fig.~\ref{fig:nev200} that the two
neutrino mass orderings lead to a quite distinct dependence of
$N^S_{ee}$,  $N^S_{\mu\mu}$, and $N^S_{e\mu}$ with the Majorana phase $\alpha$.
Since the SM background  is rather small compared to the expected
signal, we might be able to determine the neutrino ordering by simply
comparing the three different number of events for basically all values
of $\alpha$ as well as to obtain information on  the value of $\alpha$.
We will go back to this point at the end of next  section.

\FIGURE[!t]{
\includegraphics[width=0.75\textwidth]{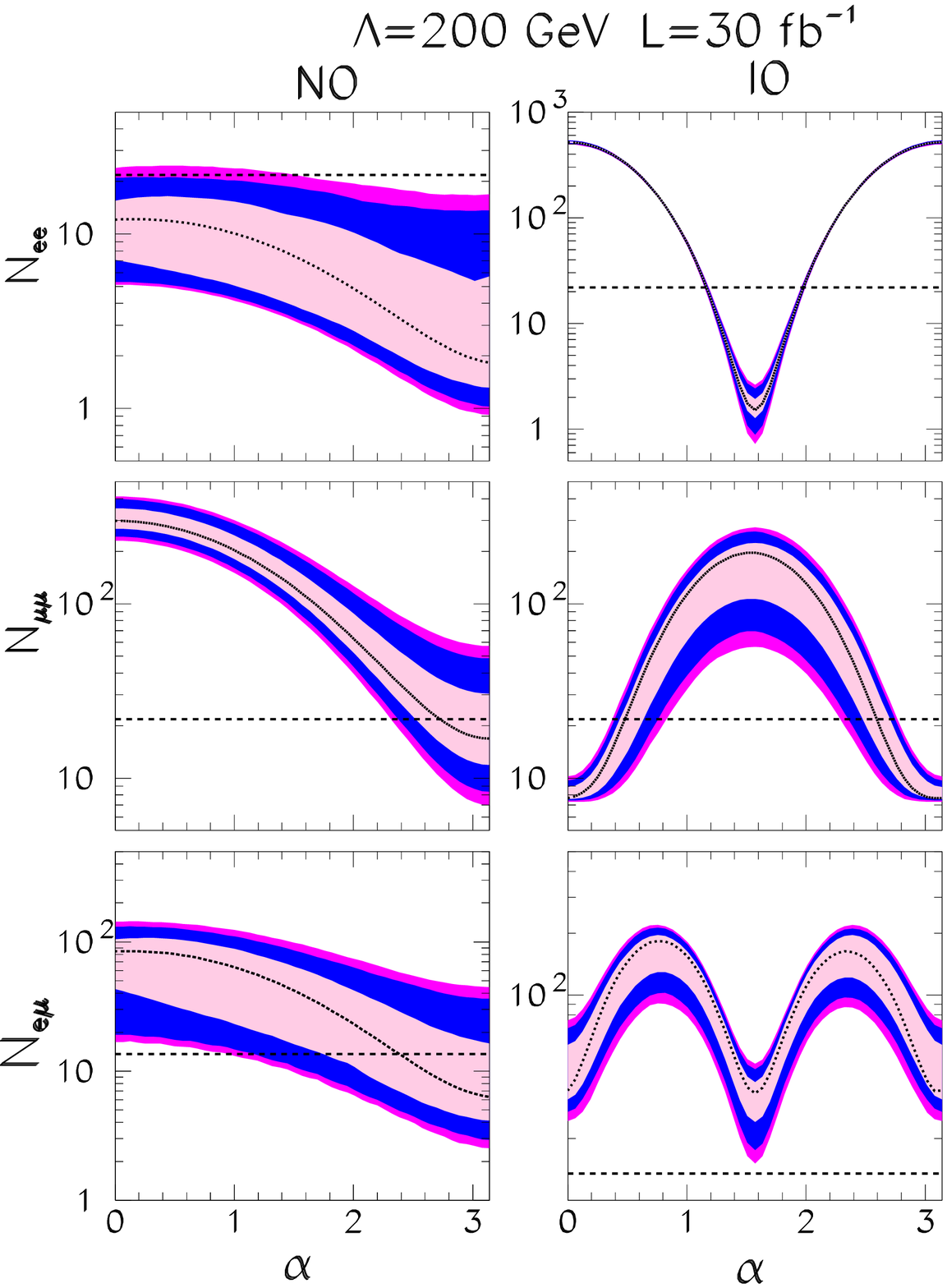}
\caption{Predicted number of events $N_{ab}$ for $pp\rightarrow
  \ell^\pm \ell^\mp \ell^\mp j j \Sla{E}_T$ with $\ell=e,\mu$ for a
  triplet fermion of mass $\Lambda=200$ GeV with an integrated
  luminosity of ${\cal L}=30$ fb$^{-1}$. 
  The horizontal dashed lines are the corresponding number of
  background events; see Table~\ref{tab:bck1}. The conventions are the
  same as in Fig.~\ref{fig:yuk}.}
\label{fig:nev200}}

Next we study the observability of this MLFV model as a function of
the triplet fermion mass $\Lambda$, the range of the neutrino
parameters $\vec\theta$ and the Majorana phase $\alpha$. We estimate the significance of the signal 
by constructing
a simple $\chi^2$ function in terms of the 
three signal  and background flavour rates for a given value of
$\Lambda$
\begin{eqnarray}
\chi^2(\vec\theta,\alpha)&=&
\sum_{ab=ee,e\mu,\mu\mu} \chi^2_{ab} (\vec\theta,\alpha)
\;,
\\\nonumber 
\chi^2_{ab}(\vec\theta,\alpha)&=&\frac{{N^S_{ab}}^2}{N^B_{ab}} 
\;\;\; \;\;\; \;\;\; \;\;\; \;\;\; \;\;\; 
\;\;\; \;\;\; \;\;\; \;\;\; \;\;\; \;\;\; \;\;\; 
{\rm For} \, N^B_{ab}\ge 10 
\; ,
\\\nonumber 
\chi^2_{ab}(\vec\theta,\alpha)&=&2(
N^S_{ab}+N^B_{ab}\ln\frac{N^B_{ab}}{N^B_{ab}+N^S_{ab}})\;\;\; \;\;\; \;\; 
{\rm For} \, N^B_{ab}< 10
\;.
\label{eq:chi}
\end{eqnarray}
Figure \ref{fig:sigt}
shows the significance --  
estimated as  $\#\sigma =\sqrt{\chi^2}$ --  of the excess of signal events 
$pp\rightarrow \ell^\pm \ell^\mp
\ell^\mp j j \Sla{E}_T$ for three values of the mass $\Lambda$ and for
${\cal L}=30$ fb$^{-1}$ as a function of the Majorana phase $\alpha$.
The significance is shown as obtained for the best fit values of
$\vec\theta$ (dashed line) and 1$\sigma$, $2\sigma$,
  and 99\% CL ranges of $\vec\theta$ (filled areas) obtained from
  the global analysis of neutrino data~\cite{ourfit}. The left
  (right) panels correspond to normal (inverted) ordering.

  As long as the number of background events is large enough, the
  results for other luminosities can be simply obtained by rescaling
  figure \ref{fig:sigt} by a factor $1/\sqrt{{\cal L}/30}$. From this
  figure we see that with ${\cal L}=100$ fb$^{-1}$ LHC can
  discover/discard this MLFV model using this channel in most of the
  presently allowed neutrino parameter space for $\Lambda\leq 300$
  GeV. In some parts of the neutrino parameter space, and in
  particular if the neutrino masses have inverse ordering, the reach
  can be extended to higher masses or to a $pp$ center--of--mass
  energy $\sqrt{s}=7$ TeV as we will discuss in Sec.\ref{sec:7tev}.

\FIGURE[!t]{
\includegraphics[width=0.75\textwidth]{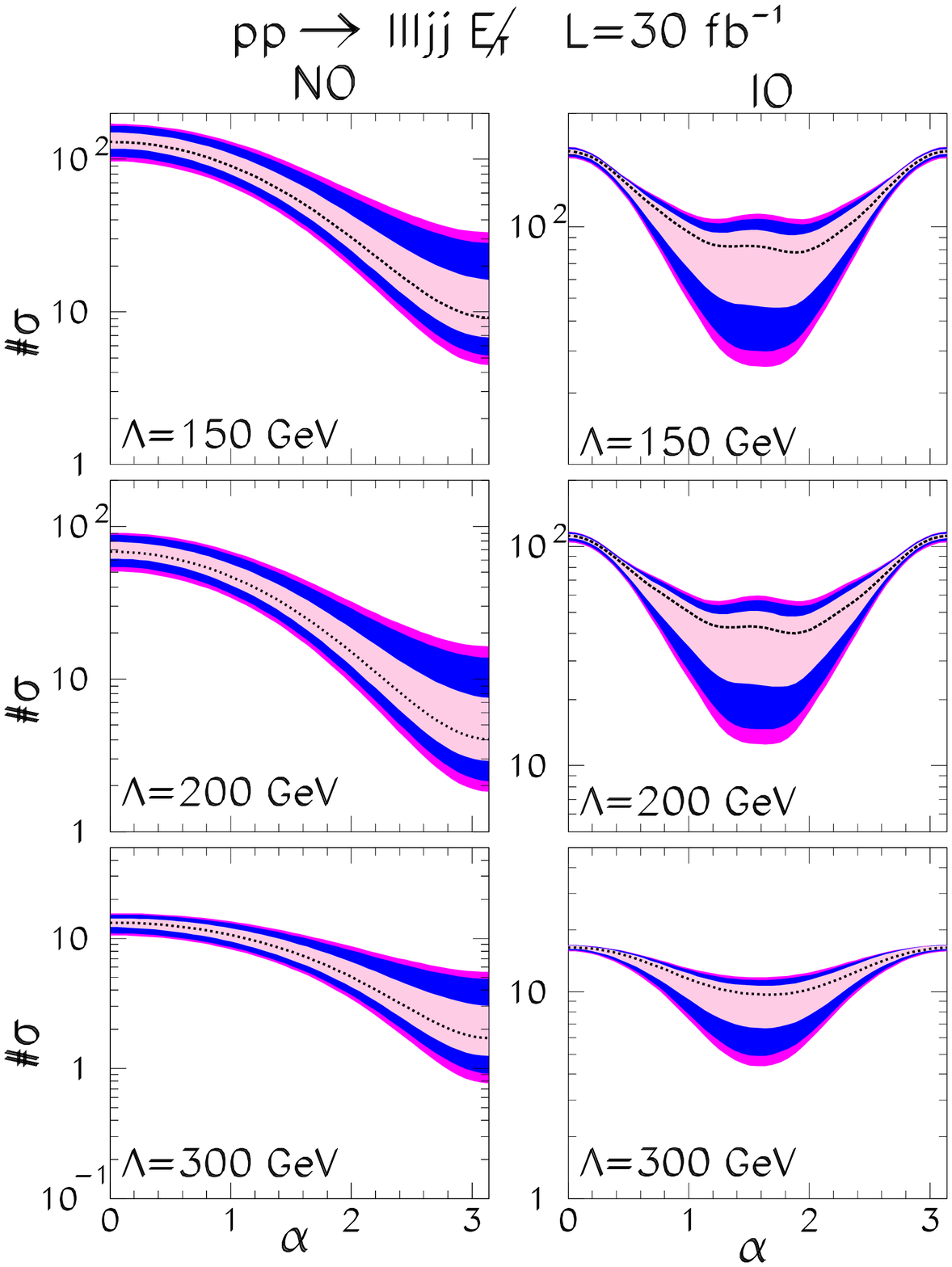}
\caption{Signal significance ($\#\sigma$) in the channel
  $pp\rightarrow \ell^\pm \ell^\mp \ell^\mp j j \Sla{E}_T$ with $\ell
  = e,\mu$ for three triplet fermion masses and for an integrated
  luminosity of ${\cal L}=30$ fb$^{-1}$ as function of the 
Majorana phase $\alpha$. 
 The conventions are the  same as in Fig.~\ref{fig:yuk}.}
\label{fig:sigt}}

\section{Process $pp\rightarrow\ell\ell jjjj$}
\label{sec:lljjjj}

The search for Type-III see--saw leptons via process \eqref{eq:proc2},
{\em i.e.}
\begin{equation}
pp \rightarrow \ell_1^\mp \,\ell_2^\pm \,j \, j \, j \, j 
\end{equation}
with $\ell_{1(2)}=e,\mu$ does not present ambiguities in the flavour
tagging, what favors the test of the MLFV hypothesis. However, 
it is plagued with a large SM background. The dominant
backgrounds for this process are:
\begin{itemize}

\item $t\bar t j j $ production where the two b's from the $t
  \rightarrow W b$ decays are identified as jets and both $W$'s decay
  leptonically;

\item $Z^*/\gamma^* jjjj$ 
with the $Z^*/\gamma^*$ leading to a charged
  lepton pair. Notice that this process only contributes to the final
  state with equal flavour leptons.

\end{itemize}
Additional backgrounds include $t\bar t W$ and $t\bar t Z$ but after
the reconstruction requirements they are very much suppressed. For further
details see Refs.~\cite{delAguila:2008cj,strumia} for a detailed
analysis of the backgrounds for this signature\footnote{ We employ
  the latest version of MadEvent MG5 \cite{MG5} in the evaluation of these
  backgrounds. For the $Z^*/\gamma^* jjjj$ process MG5 gives a 20-30\% larger
  value of this background after all the cuts are imposed as compared
  to previous versions of MadEvent.}.

Our analysis starts by applying the acceptance and isolation cuts for
the final leptons and jets, as well as the minimum transverse momentum
requirement as described in Eq.~\eqref{eq:basiccuts}. Since the signal
does not contain any undetectable particle we further required a
maximum amount of missing energy in the event
\begin{equation}
   \Sla{E}_T < 30 \hbox{ GeV}\; .
\label{eq:missmax}
\end{equation}

In what respects the reconstruction of the triplet fermions, there are
six possible ways of grouping the leptons and jets in the final state
in two sets of one lepton and two jets. We impose that at least one of
the six combinations has the two invariant masses inside the triplet
fermion mass region
\begin{equation}
 \Lambda-40 \ \mbox{GeV}  < M_{\ell jj} < \Lambda+40 \ \mbox{GeV.}
\label{eq:mrec2}
\end{equation}
Furthermore the corresponding invariant masses of the two jet pairs
are required to verify
\begin{equation}
 M_W-10\;\hbox{ GeV} < M_{jj} < m_{h^0}+10\; \hbox{ GeV.}
\label{eq:mjjwh}
\end{equation}
Since there is no ambiguity in the assignment of the two charged
leptons the total cross section of process \eqref{eq:proc2} is simply given by
\begin{equation}
\sigma^S_{ab}=\sigma_4(2-\delta_{ab})|\tilde{Y_a}|^2|\tilde{Y_b}|^2\;.
\label{eq:signajj}
\end{equation}

Even after the reconstruction of the invariant masses of the two
triplet fermions, the SM backgrounds are still large, in particular
the one arising from $Z^*/\gamma^* jjjj$. To further reduce this
background we make use of the fact that in the signal the
characteristic invariant mass of the two leptons is larger than for
the background as illustrated in Fig.~\ref{fig:mll}.  Consequently,
this background is reduced by factor $20$--$8$ for
$\Lambda=150$--$500$ GeV if we impose that the invariant mass of the
two charged leptons verify
\begin{equation}
    M_{\ell^+\ell^-}>100 \hbox{ GeV}.
\label{eq:mllmin}
\end{equation}

\FIGURE[!t]{
\includegraphics[width=3.5in]{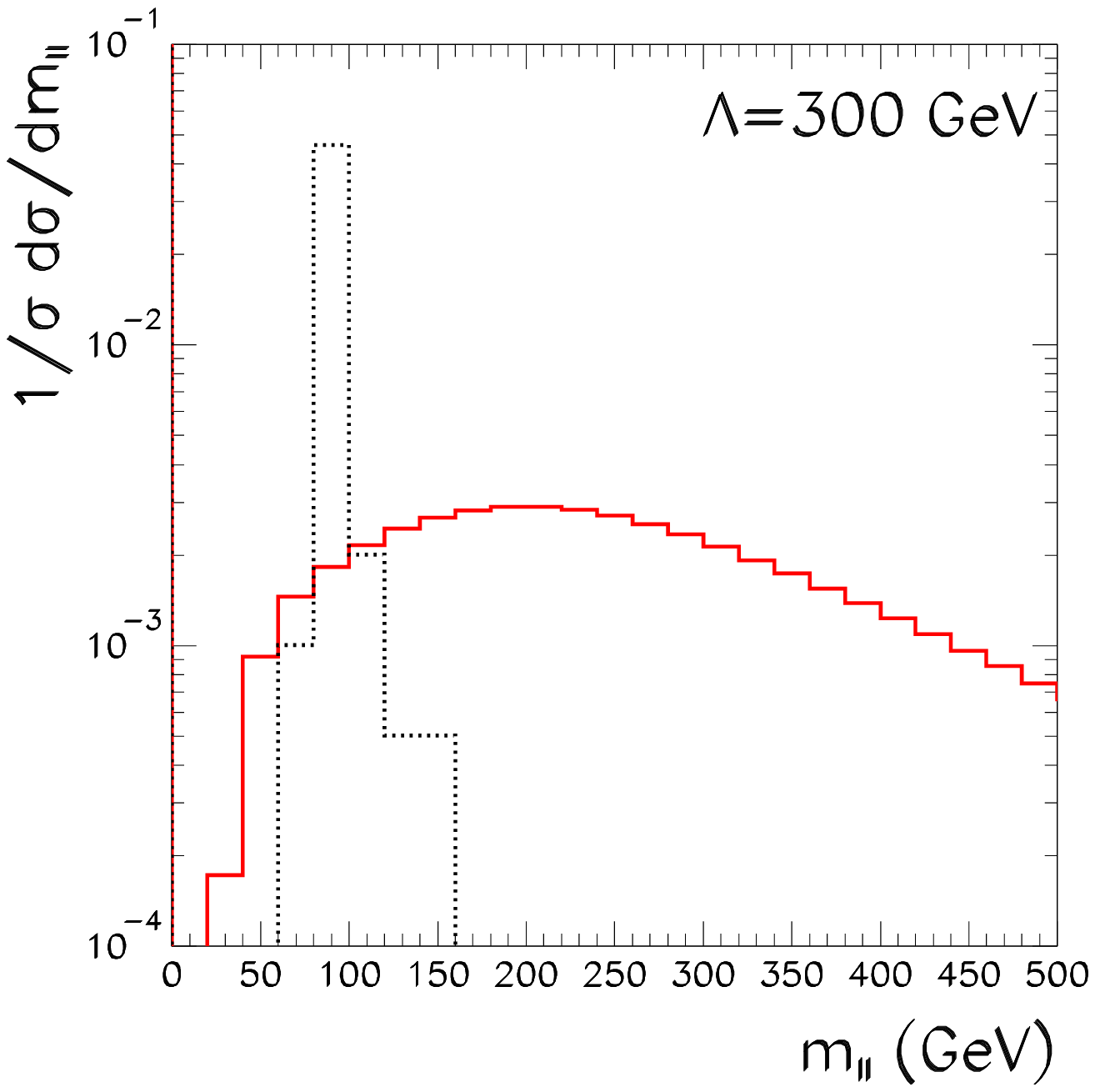}
\caption{Distribution of the charged lepton pair invariant mass for
  the signal (solid red histogram) and $Z^*/\gamma^* jjjj$ background
  (dotted black histogram) after cuts
  \eqref{eq:basiccuts},\eqref{eq:missmax}, \eqref{eq:mrec2} and
  \eqref{eq:mjjwh}, and for a heavy fermion mass $\Lambda=300$ GeV.}
\label{fig:mll}}

We present in Table \ref{tab:lljjjj} the cross sections for signal and
SM backgrounds after cuts \eqref{eq:basiccuts}, \eqref{eq:missmax},
\eqref{eq:mrec2}, \eqref{eq:mjjwh}, and \eqref{eq:mllmin}.  The
predicted number of events for the triplet fermion signal in this
channel for the different flavour combinations can be easily obtained
from Eq.~\eqref{eq:nev} and Table \ref{tab:lljjjj} using the values of
the Yukawa couplings in \eqref{eq:no1}--\eqref{eq:io3} and a detection
efficiency of $\epsilon={e^j}^4\times {e^\ell}^2=0.26$.

\TABLE[!ht]{
\centering
\begin{tabular}{||c|c|c|c||}\hline
 & Signal (fb) & \multicolumn{2}{|c||}{Background (fb)}\\ \hline
&                & $t\bar t jj$ & $Z^*/\gamma^* jjjj$ \\ \hline
$\Lambda$(GeV) & $\sigma_4$ & $\sigma^B_{ee}=\sigma^B_{\mu\mu}=
\sigma^B_{e\mu}/2.$
&$\sigma^B_{ee}= \sigma^b_{\mu\mu}$ \\
\hline
150 & 276.0 & 6.0 & 29.3\\ \hline
200 & 216.0 & 9.7 & 33.2\\ \hline
300 & 74.9 & 0.89 & 4.6\\ \hline
500 & 11.3 & 0.018 & 0.057 \\ \hline
 \end{tabular}
\label{tab:lljjjj} 
\caption{Signal and background cross sections for $pp \rightarrow \ell_a^\mp 
  \,\ell_b^\pm \,j \, j \, j \, j$ after 
  cuts \eqref{eq:basiccuts},\eqref{eq:missmax}, 
  \eqref{eq:mrec2}, \eqref{eq:mjjwh}, and \eqref{eq:mllmin} 
  for different  values of the triplet fermion mass $\Lambda$. These results
  do not include detection efficiencies for leptons and jets.}
}
\FIGURE[!t]{
\includegraphics[width=0.75\textwidth]{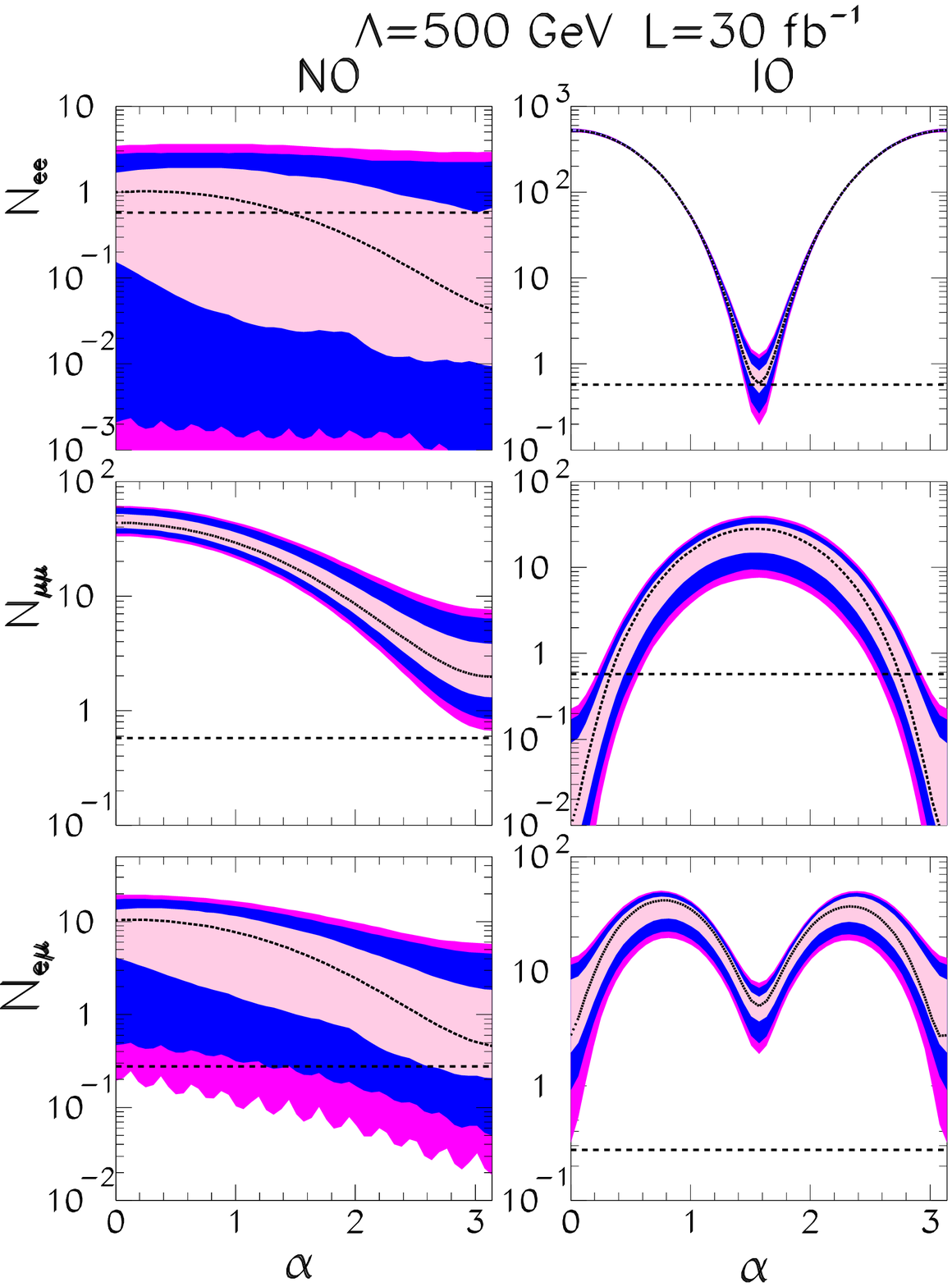}
\caption{Predicted number of events $N_{ab}$ for $pp\rightarrow
  \ell^\pm_a \ell^\mp_b j j j j $ for a triplet fermion of mass
  $\Lambda=500$ GeV with an integrated luminosity of ${\cal L}=30$
  fb$^{-1}$.
  The horizontal dashed lines are the corresponding number of
  background events.  The conventions are the same as in
  Fig.~\ref{fig:yuk}.}
\label{fig:nev500}}

We plot in Fig.~\ref{fig:nev500} the range of the expected number of
events in the different flavour combinations as a function of the
unknown Majorana phase $\alpha$ for a triplet fermion of mass
$\Lambda=500$ GeV and an integrated luminosity of ${\cal L}=30$
fb$^{-1}$.  The ranges are shown at 1$\sigma$, $2\sigma$, and 99\% CL
from the global analysis of neutrino data~\cite{ourfit}, while the
dotted line corresponds to the best fit values. The left (right)
panels correspond to normal (inverted) ordering while the horizontal
dashed lines stand for the predicted number of SM background events.
Here again, we can see that the dependence of $N^S_{ee}$, $N^S_{\mu\mu}$,
and $N^S_{e\mu}$ on the CP violating Majorana phase $\alpha$ are quite
distinct for NO and IO.

The observability of this MLFV model in the $\ell\ell j j j j$ 
channel
is depicted in Fig.~\ref{fig:sigt_4j} where we show the signal
significance as a function of the Majorana phase $\alpha$ for
different CL of the neutrino parameters. Like the previous analysis,
we added the flavours combinations to define the signal significance.  
Comparing Figs.~\ref{fig:sigt} and \ref{fig:sigt_4j} we can see that after the 
background reduction achieved by the mass reconstruction conditions
Eqs.\eqref{eq:mrec2} and \eqref{eq:mjjwh} and the 
lepton pair invariant mass cut  \eqref{eq:mllmin},  the channel
$\ell \ell jjjj$ offers better potential statistical sensitivity for the 
discovery or exclusion of this MLFV model in particular for heavier masses
$\Lambda$, despite its still larger SM backgrounds.  
One must keep in mind, however, that the final attainable  precision 
depends on the systematic  background uncertainties which are expected to 
be larger for this channel ~\cite{delAguila:2008cj}. 

\FIGURE[!t]{
\includegraphics[width=0.75\textwidth]{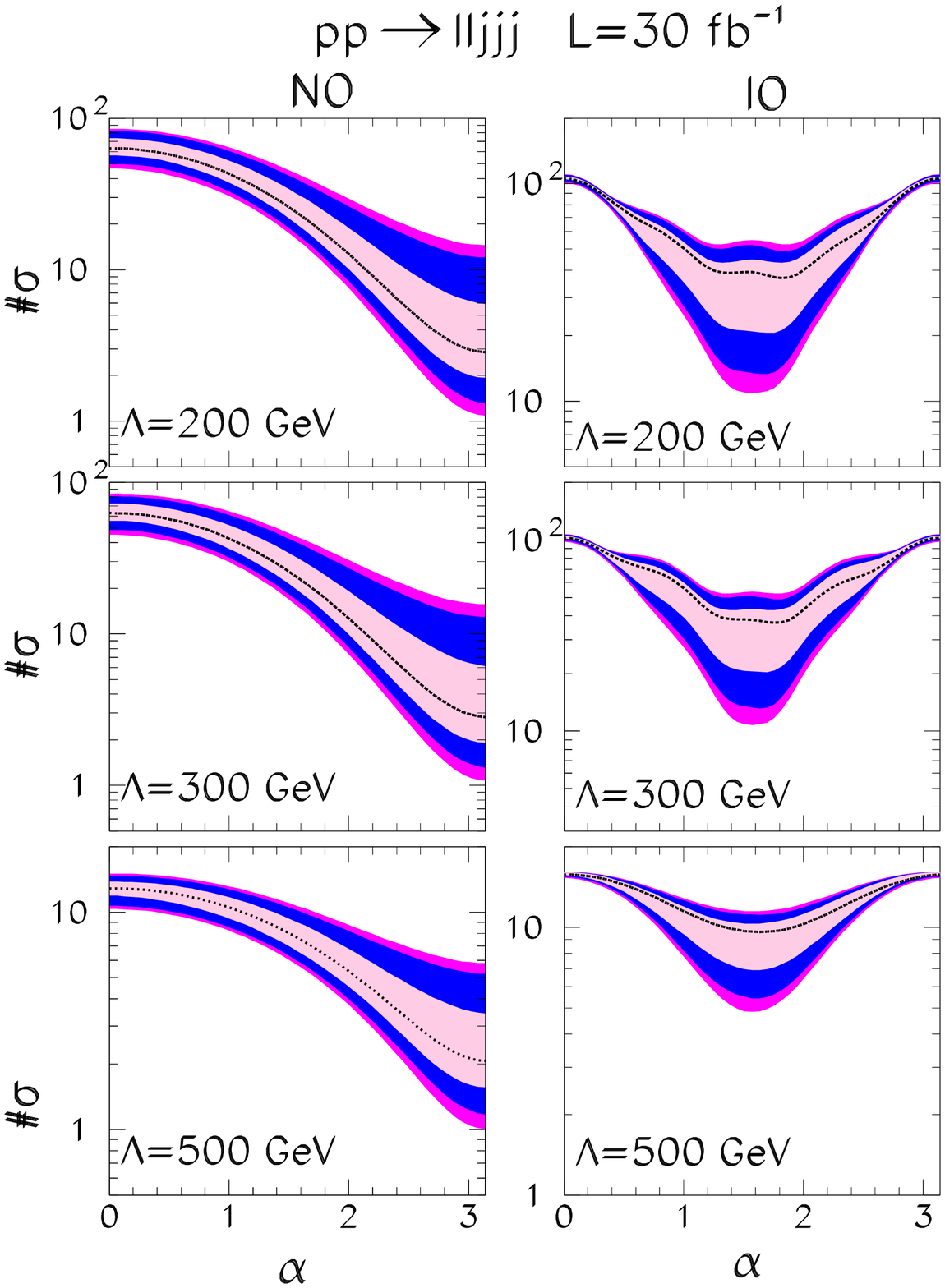}
\caption{Expected significance $\#\sigma$ of signal versus background
  events for $pp\rightarrow \ell^\pm_a \ell^\mp_b jj j j $ for three
  triplet fermion masses and for an integrated luminosity of ${\cal
    L}=30$ fb$^{-1}$. 
  The conventions are the same as in Fig.~\ref{fig:yuk}.}
\label{fig:sigt_4j}}

We can see from Figs.~\ref{fig:nev200} and ~\ref{fig:nev500} that the
two neutrino orderings lead to a very distinct dependence of
$N^S_{ee}$, $N^S_{\mu\mu}$, and $N^S_{e\mu}$ as a function of $\alpha$
for both final states. It is particularly striking the upper right
panels that presents a very narrow range for the $ee$ flavour
combination for inverted ordering and a fixed value of $\alpha$.  Thus
one expects to be able to discriminate between the inverted and normal
ordering of the neutrino masses studying the correlations between the
different flavour combinations for a large fraction of the values of
the unknown phase $\alpha$, or even to determine its value.

To illustrate this point we have assumed 
that the observed
number of events in the three flavour combinations for both
$pp\rightarrow\ell\ell\ell jj \Sla{E_T}$ and $pp\rightarrow\ell\ell
jjjj$ are those predicted for a given mass $\Lambda$ 
(assumed to be independently determined) in the NO for the best value 
of oscillation
parameters $\bar\theta_{\rm b}$ and for some fix value of the Majorana
phase $\bar\alpha$ plus the expected background events, {\em i.e.}
$N_i^{\rm obs} (\bar\theta_{\rm b},\bar\alpha)
=N^S_{i}(\bar\theta_{\rm b},\bar\alpha)+N^B_i$ with $i=1,6$
corresponding to $ee,e\mu,\mu\mu$ for the two processes 
.  We then try
reconstruct the ordering and value of $\bar \alpha$ by fitting those
six rates $N_i^{\rm obs}$ in either NO or IO with different values of
$\vec\theta$ (within their 95\% CL allowed region from oscillations)
and $\alpha$.  In order to do so we define
\begin{eqnarray}
\chi^2_{min}(\alpha)
&=&\begin{array}{l}{\rm min}\\[-0.15cm]
{\vec\theta\in 95\% {\rm CL}}\end{array}
\sum_{i=1,6 } \chi^2_i(\vec\theta,\alpha)
\\ \nonumber 
\chi^2_{i}(\vec\theta,\alpha)&=&\frac{
\left[
N^S_i(\vec\theta,\alpha)+N^B_i
-N_i^{\rm obs}(\bar\theta_{\rm b},\bar\alpha)
\right]^2}
{N_i^{\rm obs}(\bar\theta_{\rm b},\bar\alpha)}
\;\;\; \;\;\; \;\;\; \;\;\; \;\;\; \;\;\; 
{\rm for \, N_i^{\rm obs} (\bar\theta_{\rm b},\bar\alpha)
\ge 10}
\; ,
 \\\nonumber 
\chi^2_{i}(\vec\theta,\alpha)&=&2\left[
N^S_i(\vec\theta,\alpha)+N^B_i
-N_i^{\rm obs}(\bar\theta_{\rm b},\bar\alpha) \right.\\ \nonumber
&&
\left.+N_i^{\rm obs}(\bar\theta_{\rm b},\bar\alpha)
\ln\frac{N_i^{\rm obs}(\bar\theta_{\rm b},\bar\alpha)}
{N^S_i(\vec\theta,\alpha)+N^B_i}\right]
\;\;\; \;\;\; \;\;\; \;\;\; \;\;\; \;\;\; \;\;\; \;\;\; 
{\rm for \, N_i^{\rm obs} (\bar\theta_{\rm b},\bar\alpha)
< 10 \; .} 
\label{eq:chimin}
\end{eqnarray}

We plot in Fig.~\ref{fig:chimin} $\chi^2_{\rm min} (\alpha)$ for three
values of $\bar\alpha=0$, $\frac{\pi}{2}$, and $\pi$. Clearly for the
panels in the left which corresponds to the NO $\chi^2_{\rm min}
(\alpha)$ presents a minimum for $\alpha=\bar\alpha$. The panels on
the right show for which cases the event rates simulated could also be
predicted by IO with a somewhat different value of $\alpha$.  Whenever
one of the curves do not appear in the right panels it is because the
corresponding $\chi^2_{\rm min}(\alpha) >20$ for all values of
$\alpha$.

Figure~\ref{fig:chimin} illustrates that for masses $\Lambda \lesssim
200$ GeV it is possible to discriminate between NO and IO except for
$\bar\alpha\sim \frac{\pi}{2}$.  Furthermore in those cases for which
discrimination between NO and IO is possible one also obtains
information on the value of $\bar\alpha$.  As the mass increases it
becomes harder to disentangle IO and NO, and for 500 GeV for any value
of simulated $\bar\alpha$ there is always a value of $\alpha$ for
which the expected rates in IO mimic the simulated ones in NO at
better than $\sim$ 2$\sigma$.

\FIGURE[!t]{
\includegraphics[width=0.75\textwidth]{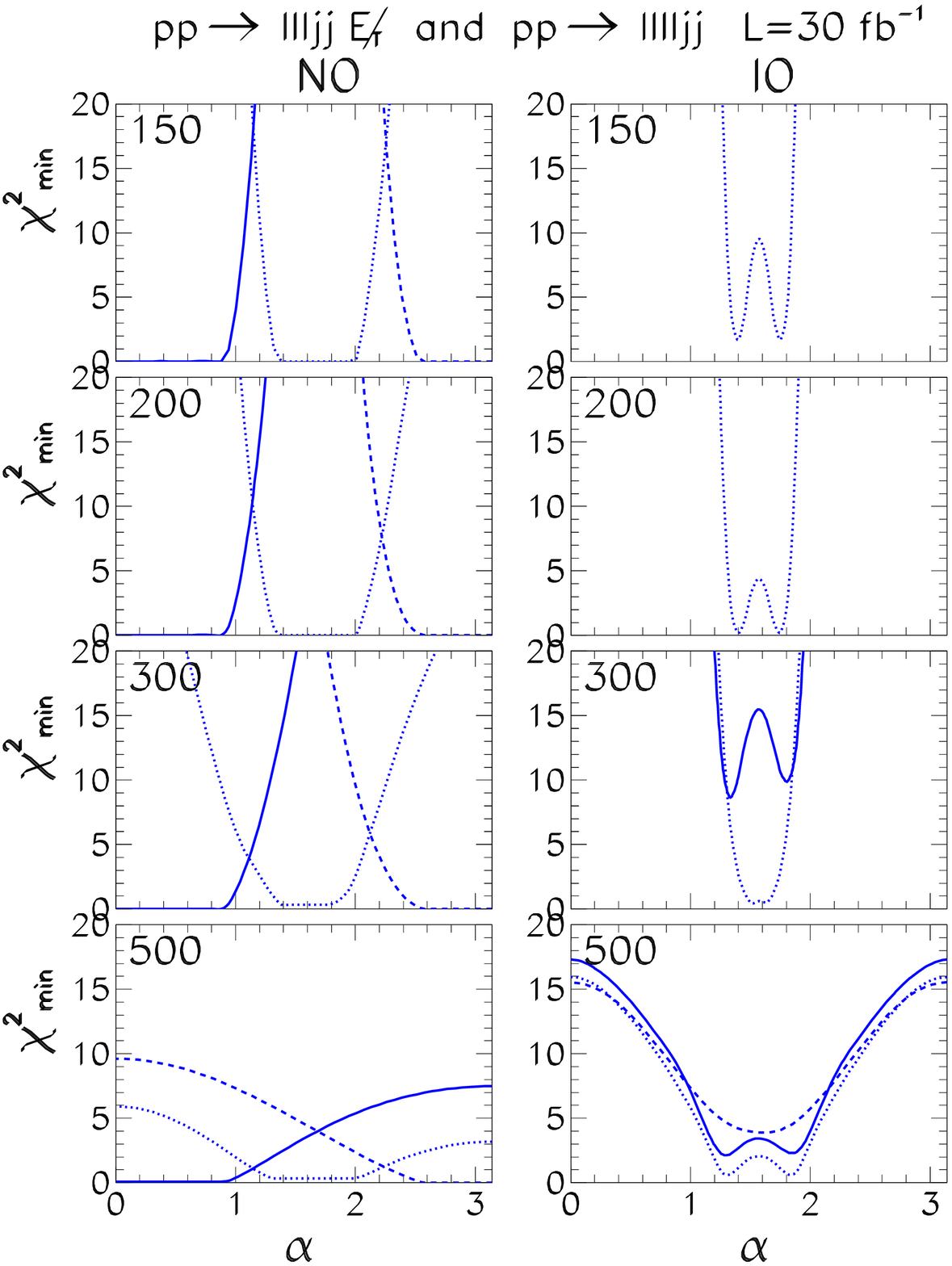}
\caption{$\chi^2_{\rm min} (\alpha)$ defined in Eq.~\eqref{eq:chimin}.
The full, dotted and dashed lines correspond to a simulated value of
the event rates  in NO for best fit values of oscillation parameters 
and  $\bar\alpha=0,\frac{\pi}{2},\pi$ respectively. 
Whenever one of the curves do not appear in the right panels it
is because the corresponding $\chi^2_{\rm min}(\alpha) >20$ for all 
values of $\alpha$.}
\label{fig:chimin}}

\section{Signals at 7 TeV}
\label{sec:7tev}

The present LHC run with a center-of-mass energy of 7 TeV has been
very successful exhibiting a rapidly increasing integrated luminosity.
Therefore, we present in this section the prospective LHC reach for
this run.  We analyzed the $pp \to \ell\ell\ell jj \Sla{E}_T$ signal at 7
TeV applying the cuts defined in
Eqs.~\eqref{eq:basiccuts}--\eqref{eq:mjjzh}, as well as, our
reconstruction procedure described in Sect.~\ref{sec:llljjn}.  We
display in Table~\ref{tab7:sig1} the contributions to the signal cross
sections ($\sigma^S_{ab}$) for the processes $pp \rightarrow \ell^\mp
\,\ell^\pm \,\ell^\pm\, j \, j \, \Sla{E_T}$ for $\ell=e,\mu$
according to Eq.~\eqref{eq:signall}. As we can see, the reconstruction
efficiency and the misidentification are at the approximately same
level for the 7 and 14 TeV runs. Of course, the signal cross section
at 7 TeV is a factor 2--4 smaller than at 14 TeV.

\TABLE[!ht]{
\centering
\begin{tabular}{||c||c|c|c|c||c||}\hline
& \multicolumn{4}{c||}{$pp \rightarrow \ell^\mp \,\ell^\pm
\,\ell^\pm\, j \, j \, \Sla{E_T}$ }& $pp \rightarrow \ell_a^\mp 
\,\ell_b^\pm \,j \, j \, j \, j$ \\\hline
$\Lambda$(\ GeV) & $\sigma_0$ & $\sigma_1$ & $\sigma_2$ & $\sigma_3$ & $\sigma_4$\\ \hline
150 & 33.2 & 0.958 & 3.25  & 0.820 & 111 \\ \hline
200 & 16.3 & 0.176 & 0.852 & 0.259 & 78.4 \\ \hline
300 & 3.72 & 0.009 & 0.036 & 0.013 & 21.0\\ \hline
\end{tabular}
\caption{Contributions to the signal cross sections ($\sigma^S_{ab}$)
in fb for the processes 
$pp \rightarrow \ell^\mp \,\ell^\pm
\,\ell^\pm\, j \, j \, \Sla{E_T}$ 
and 
$pp \rightarrow \ell_a^\mp 
\,\ell_b^\pm \,j \, j \, j \, j$ 
for $\ell=e,\mu$ according to
Eq.~\eqref{eq:signall} and Eq.~\eqref{eq:signajj} for a 7 TeV center-of-mass energy. These
results do not include the detection efficiencies for the leptons
and jets.}
\label{tab7:sig1}}

We present in Table~\ref{tab7:bck1} the main irreducible backgrounds
for the $\ell\ell\ell jj \Sla{E}_T$ channel after cuts and our
reconstruction procedure but without including detection efficiencies.
As we can see the most severe background is again the $WZjj$
production with cross sections of the order of 0.5 fb or smaller
depending on $\Lambda$. Therefore, for integrated luminosities of the
order of 10 fb$^{-1}$ we can anticipate a handful of background and
signal events for light triplet fermions, {\em i.e.} $\Lambda \lsim
200$ GeV.

\TABLE[!ht]{
\centering
 \begin{tabular}{||c|c|c|c|c|c|c||}\hline
& 
\multicolumn{2}{c|}  {$\Lambda=150$GeV}& 
\multicolumn{2}{c|}   {$\Lambda=200$GeV}& 
\multicolumn{2}{c||}  {$\Lambda=300$GeV} \\ \hline
 Process & 
$\sigma^B_{ee}=\sigma^B_{\mu\mu}$ & $\sigma^B_{e\mu}$&
$\sigma^B_{ee}=\sigma^B_{\mu\mu}$ & $\sigma^B_{e\mu}$&
$\sigma^B_{ee}=\sigma^B_{\mu\mu}$ & $\sigma^B_{e\mu}$\\ \hline
 $pp\rightarrow t\bar{t}W$ & 0.0073 & 0.0129
&0.0074 & 0.0147
&0.0010  & 0.0017
\\ \hline
 $pp\rightarrow t\bar{t}Z$ & 0.0187 & 0.0143
& 0.0214 & 0.0161 
& 0.0057 & 0.0017
\\ \hline
 $pp\rightarrow WZjj$ & 0.5653 & 0.3656
& 0.5475 & 0.3053
& 0.0709 & 0.0336
\\ \hline
 $pp\rightarrow ZZjj$ & 0.0138 & 0.0107
& 0.0133 & 0.0082
  & 0.0012 & 0.0007
\\ \hline
  Total $(\ell\ell\ell jj\nu)$ & 0.605 & 0.403  
& 0.590 & 0.344 
& 0.079  & 0.038
\\ \hline
 $pp\rightarrow t\bar{t}jj$ & 1.30 & 2.60
& 1.93 & 3.86 
& 0.15 & 0.30
\\ \hline
$pp\rightarrow Z^*/\gamma^* jjjj$ & 10.3 & 0
& 11.3 & 0 
& 0.26 & 0
\\ \hline
  Total $(\ell\ell jjjj)$& 11.60 & 2.60  
& 13.23 & 3.86 
& 0.41  & 0.30
\\ \hline
 \end{tabular}
\caption{SM background cross sections ($\sigma^B_{ab}$)
in fb for the processes $pp \rightarrow \ell^\mp \,\ell^\pm
\,\ell^\pm\, j \, j \, \Sla{E_T}$ and $pp \rightarrow \ell_a^\mp 
\,\ell_b^\pm \,j \, j \, j \, j$ for $\ell=e,\mu$ and a 7 TeV center-of-mass energy.
The results are presented for different values of the triplet
fermion mass $\Lambda$ and they do not include the detection
efficiencies for the leptons and jets.}
\label{tab7:bck1}}

We also analyzed the $\ell \ell jjjj$ channel at 7 TeV applying the
cuts and reconstruction procedure stated in Eqs.~\eqref{eq:basiccuts}
and \eqref{eq:missmax}--\eqref{eq:mjjwh}. The signal and background 
cross sections after cuts
without including the detection efficiencies are presented in Tables
\ref{tab7:sig1} and \ref{tab7:bck1}. Once more, the most
relevant background for equal flavour leptons is $Z^*/\gamma^* jjjj$
production while the $t \bar{t} jj$ process is dominant for different
flavour leptons.

Figure~\ref{fig:res7} depicts the significance of the signal versus
background for different values of the heavy fermion mass and assuming
the same detection efficiencies than for 14 TeV and an integrated
luminosity of 10 fb$^{-1}$.  To compensate for the smaller statistics
the event rates for the three flavour channels have been added for
each of the two processes and the significances for two total event
rates have been combined.  As before, the predictions are shown as
obtained for the best fit (dashed line) and 1$\sigma$, $2\sigma$, and
99\% CL neutrino parameter ranges (filled areas) obtained from the
global analysis of neutrino data \cite{ourfit}.  We can see from this
figure that integrated luminosities of the order of 10 fb$^{-1}$ can
lead to the discovery of the new heavy fermions in a significant range
of $\alpha$ for masses $\Lambda \lsim 200$ GeV.

\FIGURE[!t]{
\includegraphics[width=0.75\textwidth]{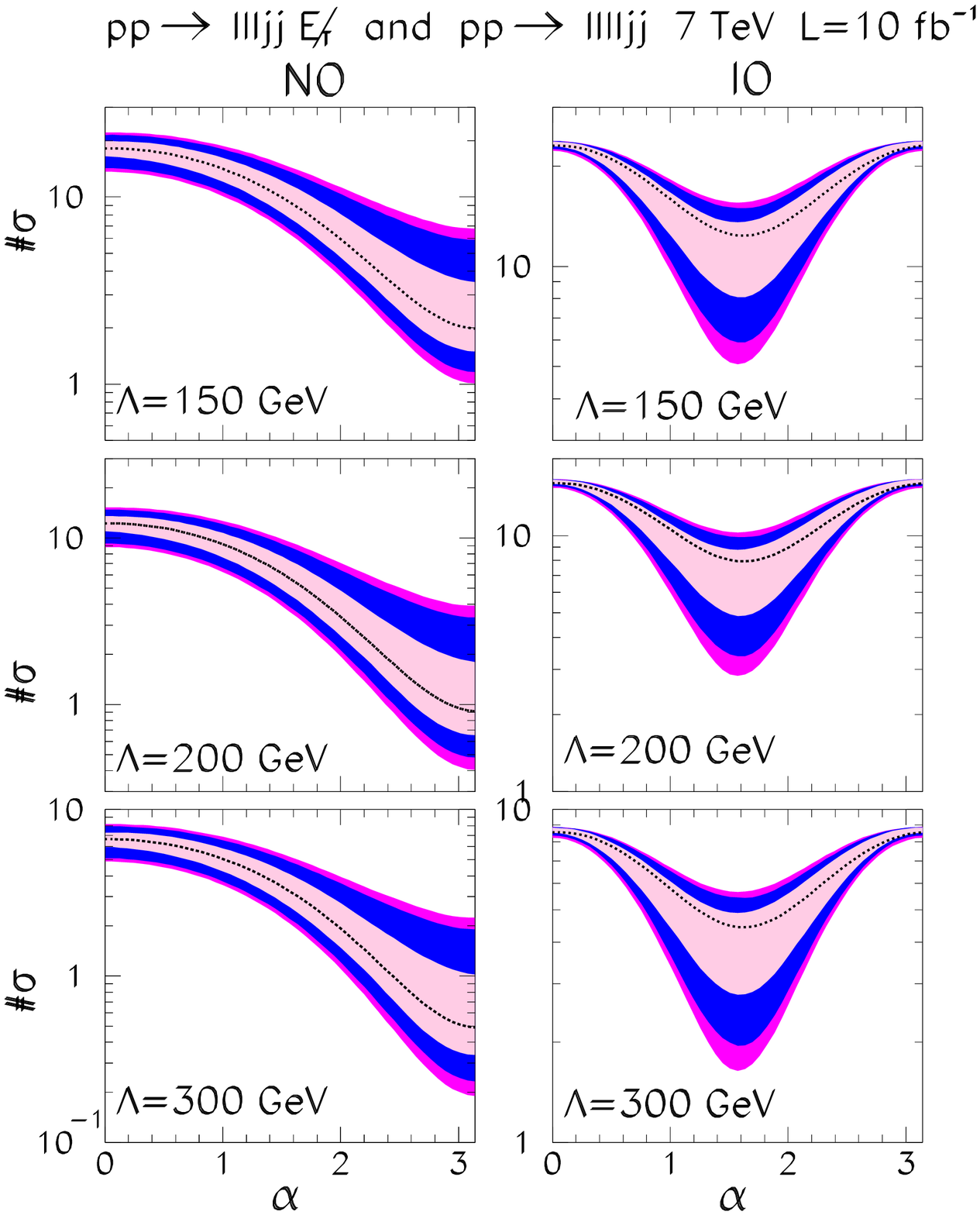}
\caption{Expected significance $\#\sigma$ of signal versus background
  events for a center--of--mass energy of 7 TeV and assuming the
  detection efficiencies used in the 14 TeV analyses.   
  The conventions are the same as in Fig.~\ref{fig:yuk}.}
\label{fig:res7}}

\section{Summary and discussions}
\label{sec:conclu}

In this work we have analyzed the signal of Type-III see--saw models with
MLFV taking into account the constraints emanating from low energy
neutrino data. We have presented our results as a function of the Majorana
phase $\alpha$ for the best fit and 1$\sigma$, $2\sigma$, and 99\% CL
neutrino parameter ranges obtained from  the global analysis of
neutrino data~\cite{ourfit}. We have optimized the analysis for a
center--of--mass energy of 14 TeV, however in Sec.\ref{sec:7tev} we
have also studied the potential of the LHC running at 7 TeV and with
integrated luminosity ${\cal O}(10)$ fb$^{-1}$ to probe part of the
parameter space of this model.

After careful analyses of the signal and SM backgrounds we have established
that mass scales of the order of 300 (500) GeV can be probed in the
channel $\ell\ell\ell jj \Sla{E}_T$ ($\ell \ell j j j j$). It is
interesting to notice that the $\ell \ell j j j j$ final state can be the
best discover channel for triplet fermions at the LHC if their 
larger SM backgrounds are well understood. 
Moreover, once a signal of Type-III
see--saw models with MLFV is observed its energy scale $\Lambda$ can
be precisely determined by measuring the mass of the new heavy
fermions; as an illustration see Fig.~\ref{fig:mrec}.  One very clean
channel for this determination is the production of the charged lepton
$E_1^\pm$ followed by its decay into three leptons, provided there
will be enough integrated luminosity available.

Finally let us comment that the discovery at the LHC of the
triplet fermions predicted by MLFV Type-III see--saw models is not only
important for unraveling the mechanism responsible for the tiny
observed neutrino masses, but it may also allow for the determination of
the ordering of the neutrino masses. In fact, the ratio of the
flavour combinations $ee$, $\mu\mu$, and $e\mu$ can discriminate
between inverted and normal ordering as we can see from
Figs.~\ref{fig:nev200}, \ref{fig:nev500} and \ref{fig:chimin}.


\acknowledgments
We thank E. Nardi for comments. 
M.C. G-G thanks the CERN Theory group for their hospitality and her
collaborators in the neutrino oscillation analysis M. Maltoni and J. Salvado.
O.J.P.E is supported in part by Conselho Nacional de Desenvolvimento
Cient\'{\i}fico e Tecnol\'ogico (CNPq) and by Funda\c{c}\~ao de Amparo
\`a Pesquisa do Estado de S\~ao Paulo (FAPESP); M.C.G-G is also
supported by USA-NSF grant PHY-0653342, 
by CUR Generalitat de Catalunya grant
2009SGR502 and together with J.G-F by MICINN FPA2010-20807
and consolider-ingenio 2010 program CSD-2008-0037. J.G-F is
further supported by Spanish ME FPU grant AP2009-2546.


\end{document}